\documentclass[useAMS,usenatbib]{mn2e}

\usepackage{amssymb}
\usepackage{amsfonts}
\usepackage{mncite}
\usepackage{epsfig}
\usepackage{psfig}

\newcommand{\nablav}{\mbox{\boldmath$\nabla$}}
\newcommand{\de}{\mbox{d}}

\begin{document}

\title[Locality of transport in self-gravitating discs]{Testing the
  locality of transport in self-gravitating accretion discs}

\author[G. Lodato \& W.K.M. Rice]{G. Lodato$^1$ and W.K.M. Rice$^2$ \\
$^1$ Institute of Astronomy, Madingley Road, Cambridge, CB3 0HA\\
$^2$ School of Physics and Astronomy, University of St. Andrews, North
  Haugh, St. Andrews KY16 9SS}

\maketitle
\begin{abstract}
In this paper we examine the issue of characterising the transport
associated with gravitational instabilities in relatively cold discs,
discussing in particular the conditions under which it can be
described within a local, viscous framework.  We present the results
of global, three-dimensional, SPH simulations of self-gravitating
accretion discs, in which the disc is cooled using a simple
parametrisation for the cooling function. Our simulations show that
the disc settles in a ``self-regulated'' state, where the axisymmetric
stability parameter $Q\approx 1$ and where transport and energy
dissipation are dominated by self-gravity. We have computed the
gravitational stress tensor and compared our results with expectations
based on a local theory of transport. We find that, as long as the
disc mass is smaller than $0.25M_{\star}$ and the aspect ratio
$H/R\lesssim 0.1$, transport is determined locally, thus allowing for
a viscous treatment of the disc evolution.
\end{abstract}
\begin{keywords}
accretion, accretion discs -- gravitation -- instabilities -- stars:
formation -- galaxies: active
\end{keywords}

\section{Introduction}

One of the basic unknowns of accretion disc theory is the physical
mechanism ultimately responsible for angular momentum transport and
energy dissipation in the disc.  It is well known that classical
hydrodynamical viscosity is not sufficient to provide accretion at the
rates inferred from the observations in almost every astrophysical
context where accretion discs play a role. The usual way to overcome
this difficulty is to assume that transport is dominated by some
``anomalous'' viscous phenomenon, possibly related to collective
instabilities in the disc, and to to give some {\it ad hoc}
parameterisations for the viscosity.  The most widely used of such
parameterisation is the so-called $\alpha$-prescription \citep[see
Section 2]{shakura73}.

It has been recently recognised that accretion discs threaded by a
weak magnetic field are subject to MHD instabilities (see
\citealt{balbusreview} and references therein), that can induce
turbulence in the disc, thereby being able to transport angular
momentum and to promote the accretion process.  However, in many
astrophysically interesting cases, such as the outer regions of
protostellar discs, the ionisation level is expected to be low,
reducing significantly the effects of magnetic fields in determining
the dynamics of the disc, at least over limited radial ranges
\citep{gammie96}. A possible alternative source of transport in cold
discs is provided by gravitational instabilities
(\citealt{lin87,laughlin94,armitage2001}). Moreover, in environments
where the disc mass is a significant fraction of the central object
mass, such as during the assembly of protostellar discs, it is
inevitable that disc self-gravity will play a role.

The axisymmetric stability of a thin disc with respect to its own
self-gravity is determined by the parameter $Q$ \citep{toomre64},
defined as:
 \begin{equation}
Q=\frac{c_s\kappa}{\pi G \Sigma},
\label{Q}
\end{equation}
where $c_s$ is the effective thermal speed of the disc, $\Sigma$ is
the surface density, and $\kappa$ is the epicyclic frequency (which,
for a Keplerian disc, is equal to the angular velocity $\Omega$). The
disc is unstable if $Q$ is smaller than a threshold value
$\bar{Q}\approx 1$. It has been long recognised, especially in the
context of galactic dynamics \citep{hohl71,bertinromeo88}, that the
development of gravitational instabilities would lead to a
self-regulation process: if the disc is initially cold, in the sense
that $Q<\bar{Q}$, then gravitational instabilities would heat it up on
the fast dynamical timescale, bringing it toward stability; on the
other hand, if the disc is hot enough to begin with, then radiative
cooling is going to bring the value of $Q$ down toward an unstable
configuration.  As a result of the presence of these two competing
mechanisms, the ``switch'' associated with the onset of gravitational
instabilities will act as a thermostat for the disc, which is
therefore expected to be always close to marginal stability. A similar
approach has been also suggested in the case of accretion discs
\citep{pacinski78,bertin97}.

From the observational point of view, there are already some clues
that the disc self-gravity can be important both in the context of
protostellar discs and in accretion discs around supermassive black
holes in Active Galactic Nuclei (AGN). However, a detailed comparison
with observations is limited by the lack of detailed models of
self-gravitating discs and by an incomplete understanding of the basic
physical processes involved.

The importance of the disc self-gravity in observations of
protostellar discs was pointed out by \citet{adams88}, who realized
that a massive disc, with a flat rotation curve, could reproduce the
observed flat long-wavelength Spectral Energy Distribution (SED) of
many protostellar sources. However, in these early studies no detailed
model of self-gravitating accretion discs was available, so that the
required disc mass turned out to be unreasonably large. These ideas
were recently revisited by \citet{LB2001} in the light of a
self-consistent disc model \citep{BL99}, and they were able to model
the observed SED of FU Orionis objects (a class of young stellar
objects undergoing a phase of enhanced accretion), with a disc mass
comparable to that of the central object. This model assumes that the
outer disc (beyond a few au from the central star) is
self-regulated. However, non-self-gravitating $\alpha$-models of
accretion discs would predict a rapidly declining radial profile of
$Q$, which would eventually become unphysically small in the outer
regions of the disc.  In the case of FU Orionis discs the disc is
predicted to be marginally stable already at a distance of $\approx 1
$ au from the central star \citep{bellin94}.  The arguments at the
basis of the self-regulation mechanisms would suggest that in the
outer regions of the disc an additional heating source is required in
order to keep the disc marginally stable.  \citet{LB2001} argued that
the difficulty with $\alpha$-models arises because these viscous
models assume that energy dissipation is determined locally, whereas
gravitational instabilities would naturally act in a global way,
leading to a modification of the standard estimates of the viscous
dissipation power.

Analogous considerations also hold in the context of AGN discs. Here,
\citet{LB03a} have shown that the non-Keplerian rotation curve traced
by water masers in the Seyfert galaxy NGC 1068
\citep{greenhill96,greenhill97} can be reproduced by self-regulated
disc models. \citet{sirko03}, following the same arguments of
\citet{LB2001}, have modelled the long-wavelength SED of AGN discs,
based on the requirement that $Q\approx 1$. The latter authors also
recognise the need for some additional heating, but, contrary to
\citet{LB2001}, attribute it to some ``external'' source, namely
nuclear fusion in stars embedded in the disc (see also
\citealt{collin99}).

Actually, the issue of locality of transport in self-gravitating
accretion discs still remains an open question.  \citet{lin87} have
suggested that the transport induced by self-gravity could be
described within a viscous framework, and introduced a modified
$\alpha$-prescription, where $\alpha\sim Q^{-2}$.  In this way, a
self-regulated disc would be characterised by a rather large effective
viscosity, with $\alpha\approx 1$. On the other hand, \citet{balbus99}
have shown that the energy equation for self-gravitating discs cannot
be put in the form of a diffusion equation, as required in a viscous
scenario, and that the energy flux contains some extra terms,
associated with global wave transport, that are ``anomalous'' from the
point of view of viscosity. Similar ideas had also been suggested by
\citet{shu90}.

In this context, numerical simulations play a central role.
\citet{laughlin94} have performed global, smoothed particle
hydrodynamics (SPH) simulations of massive isothermal discs,
concluding that the density evolution of the disc could be reproduced
in terms of simple $\alpha$-models. \citet{laughlin96}, using 2D grid
based simulations, have shown that in order to reproduce the density
evolution induced by gravitational instabilities an $\alpha$
coefficient dependent on radius was needed.  However, these
simulations did not include a detailed treatment of the heating and
cooling processes in the disc, which have been shown to play a
fundamental role in determining the outcome of the instability
\citep{pickett2000,nelson2000}.

\citet{gammie01} has performed local, shearing-sheet, zero-thickness
simulations of self-gravitating discs, including a simple cooling
term, and has concluded that a local description is adequate in such
``razor-thin'' discs.  However, Gammie's simulations are not
appropriate to test global effects, since locality is set up from the
beginning, and they are only valid for infinitesimally thin discs,
while the typical distance over which gravitational instabilities are
expected to travel scales with the disc thickness.
\citet{rice03a,rice03b} have already shown using global,
three-dimensional simulations, how global effects can be important in
the dynamics of self-gravitating discs, in relation to the related
issue of the fragmentation of a massive disc.

In this paper we present the results of global, three-dimensional, SPH
simulations of massive, cooling, non-magnetised discs.  Our main
purpose is to quantitatively determine the dissipation power provided
by gravitational instabilities and to compare the results with the
expectations based on a viscous theory of discs, in order to assess
the extent to which the transport induced by gravitational
instabilities can be regarded as a local process.

The paper is organised as follows.  In Section \ref{sec:general} we
summarise the basic transport properties of viscous and of
self-gravitating discs, introducing the basic physical quantities
involved. In Section \ref{sec:setup} we describe the numerical setup
of our simulations. In Section \ref{sec:results} we show the results
of our computations.  In Section \ref{sec:discussion} we discuss our
results in comparison with previous investigations and in relation to
observed systems. In Section \ref{sec:conclusions} we draw our
conclusions. In the Appendix we discuss the more technical issue of
transport induced by artificial viscosity in the numerical simulations
presented.

\section{Transport in accretion discs}
\label{sec:general}

In this Section we will summarise a few well-known results about the
dynamics of viscous and self-gravitating accretion discs, that are
going to be essential in the description of our results. We will not
go into the full details of the derivation of the main results, for
which one can refer to standard reviews, such as that of
\citet{pringle81}.

\subsection{Non-self-gravitating discs}

In the analysis performed in this paper we will adopt the thin disc
approximation, and will therefore deal with vertically integrated
quantities. The equations of motion for an axisymmetric disc, in
cylindrical coordinates, are the equation of continuity:
\begin{equation}
\label{continuity}
\frac{\partial\Sigma}{\partial t}+\frac{1}{R}\frac{\partial}{\partial
R} (R\Sigma u) =0,
\end{equation}
and the azimuthal component of Euler's equation (expressed in terms of
angular momentum conservation):
\begin{equation}
\label{angmomcons}
\frac{\partial}{\partial t}(\Sigma
R^2\Omega)+\frac{1}{R}\frac{\partial} {\partial R} (\Sigma
R^3\Omega u)=-\frac{1}{R}\frac{\partial}{\partial R}(R^2 T_{R\phi}),
\end{equation}
where $u$ is the radial velocity and $T_{R\phi}$ is the relevant
component of the viscous stress tensor, integrated in the vertical
direction. This last term is the basic ingredient in the theory of
accretion discs.  As we have already anticipated, standard
hydrodynamical viscosity is not sufficient to provide accretion at the
required rates, and therefore $T_{R\phi}$ is generally described by
means of {\it ad hoc} prescriptions.  The $\alpha$-prescription
\citep{shakura73}, based on simple arguments on the relevant physical
scales of turbulent cells in the discs, assumes that $T_{R\phi}$ is
proportional to the disc pressure:
\begin{equation}
T_{R\phi}=\left|\frac{\mbox{d}\ln\Omega}{\mbox{d}\ln
  R}\right|\alpha\Sigma c_s^2,
\label{alpha}
\end{equation}
where the proportionality constant $\alpha$ is an unknown parameter,
usually considered to be smaller than unity. The $\alpha$-prescription
can be also put in the following equivalent form, which involves the
kinematical viscosity coefficient $\nu$:
\begin{equation}
\label{eq:alphashakura}
\nu=\alpha c_{\mathrm s} H,
\end{equation}
where $H$ is the thickness of the disc.

The effect of viscosity on the energy balance is twofold: viscous
torques convect energy between neighbouring annuli of the disc and they
dissipate energy. The energy which is convected across a ring at
radius $R$ per unit time is given by:
\begin{equation}
\label{convection}
\frac{\de E}{\de t}=2\pi R^2 T_{R\phi}\Omega,
\end{equation}
while the dissipation rate per unit surface is given by:
\begin{equation}
\label{dissipation}
D(R)=T_{R\phi} |R\Omega'|.
\end{equation}
If the disc is in thermal equilibrium, we can derive a useful relation
between the viscosity coefficient $\alpha$ and the cooling
timescale. If we assume that cooling can be simply parameterised in
the following way:
\begin{equation}
\label{cooling}
Q^-=\frac{U}{t_{\mathrm{cool}}}=\frac{\Sigma c_s^2}{\gamma(\gamma-1)
t_{\mathrm{cool}}},
\end{equation}
where $U$ is the internal energy per unit surface and $\gamma$ is the
ratio of the specific heats, then, equating the viscous dissipation
term, expressed in Eq.  (\ref{dissipation}) to the cooling term,
expressed in Eq. (\ref{cooling}), leads to:
\begin{equation}
\label{alphavisc}
\alpha=\left|\frac{\de\ln\Omega}{\de\ln R}\right|^{-2}\frac{1} 
{\gamma(\gamma-1)t_{\mathrm{cool}}\Omega},
\end{equation}
where we have used also Eq. (\ref{alpha}).

\subsection{Self-gravitating discs}
\label{sec:self}

We will now turn to the transport properties of self-gravitating
discs.  The gravitational potential due to the disc will be denoted by
$\Phi_s$, and ${\bf g}=-\nablav\Phi_s$ is the gravitational field.

It can be shown \citep{lyndenbell72} that the equation of angular
momentum conservation can be put in a form analogous to Eq.
(\ref{angmomcons}), as required in a viscous framework, where the
gravitational stress tensor is:
\begin{equation}
T_{R\phi}^{\mathrm{grav}}=\int\mbox{d}z\frac{g_{\mathrm{R}}g_{\phi}}{4\pi
G}.
\label{stress}
\end{equation}
Eq. (\ref{stress}) only accounts for the transport induced by the
gravitational field itself. Gravitational instabilities will also
produce density and velocity perturbations that contribute to the
transport and should be included in the calculations. This
contribution (the ``Reynolds'' stress)  can be expressed as:
\begin{equation}
T_{R\phi}^{\mathrm{Reyn}}=\Sigma {\delta v}_R{\delta v}_{\phi},
\label{hydro}
\end{equation}
where ${\bf{\delta v}=\bf{v}-\bf{u}}$ is the velocity fluctuation,
with $\bf{v}$ the fluid velocity and $\bf{u}$ the mean fluid
velocity.

Given an expression for the gravitational stress tensor, one could
therefore be tempted to use the $\alpha$-prescription, and to assume
that $T_{\mathrm{R\phi}}$ is simply proportional to the local
pressure. If large-scale structure and global processes play a role in
self-regulating the disc, it could even be possible to give a
``generalised'' $\alpha$-prescription, where $\alpha$ is to be
determined by some global requirement (on this point, see
\citealt{bertin97,coppi80}).  In any case, even if such {\it global}
parameterisation is possible, it should be emphasised that the
previous comments only apply to angular momentum transport. We still
have to face the issue of energy transport in self-gravitating discs.
In particular, we should check whether gravitational instabilities
transport energy between neighbouring annuli according to Eq.
(\ref{convection}) and whether they dissipate energy according to
Eq. (\ref{dissipation}).

\citet{balbus99} have shown that in general energy transport cannot be
described viscously. The energy balance equation for self-gravitating
discs contains some ``extra'' terms (see Eq. (59) in
\citealt{balbus99}), that Balbus \& Papaloizou ascribe to global wave
transport.  The important issue at this stage is to determine how
important these extra terms are, and under which conditions they are
able to influence the energy balance in the disc.

In this work we will use global numerical simulations in order to
compute explicitly the different physical quantities described in this
Section.  In particular we will evaluate the gravitational stress
tensor, according to Eqs. (\ref{stress}) and (\ref{hydro}), and the
corresponding ``viscous'' dissipation term, which can be then directly
compared to the power actually dissipated in our simulated discs.

\section{Numerical setup}
\label{sec:setup}

\subsection{The code}

The three-dimensional simulations presented here were performed using
smoothed particle hydrodynamics (SPH), a Lagrangian hydrodynamic code
(see \citealt{benz90,monaghan92}). In these simulations the central
object is modelled as a point mass onto which gas particles can
accrete if they approach to within the sink radius, while the gaseous
disc is simulated using 250,000 SPH particles.  Both the point mass
and the gas particles use a tree to determine neighbours and to
calculate gravitational forces \citep{benz90}, and the central object
is free to move under the influence of the disc gas.  A significant
saving in computational time is made by using individual, particle
time-steps \citep{bate95} with the time-steps for each particle
limited by the Courant condition and by a force condition
\citep{monaghan92}.

Since the main aim of this work is to check the energy processes
associated with gravitational instabilities, we explicitly solve the
energy balance equation for the gas. We allow the disc to heat up due
to both $P\de V$ work and viscous dissipation. The ratio of the
specific heats is assumed to be $\gamma=5/3$. For the cooling, we
follow \citet{gammie01} and add a simple cooling term to the energy
equation. Specifically, we use the same kind of parametrisation as in
Eq. (\ref{cooling}). For a particle with internal energy per unit mass
$u_i$, the cooling is implemented using
\begin{equation}
\label{cooling2}
\frac{\de u_i}{\de t}=-\frac{u_i}{t_{\mathrm{cool}}},
\end{equation} 
with $t_{\mathrm{cool}}=\beta \Omega^{-1}$. \citet{gammie01} and
\citet{rice03b} have shown that for small cooling times the disc may
fragment into gravitationally bound objects while for longer cooling
times, the disc settles into a quasi-steady state with the imposed
cooling balanced by dissipation through the growth of the
gravitational instability. In particular, \citet{gammie01} has shown
that the fragmentation boundary occurs for $t_{\mathrm{cool}} \le
3\Omega^{-1}$, while \citet{rice03b} show that global effects may lead
to an enhanced tendency for fragmentation, so that the critical value
for $t_{\mathrm{cool}}$ is somewhat increased.  The reason why a disc
that cools too rapidly is more prone to fragmentation is because
gravitational heating of the disc occurs on the dynamical timescale,
so if the disc cools very rapidly, self-gravity has not enough time to
prevent the formation of bound objects in the disc.  In our
simulations we have adopted $\beta=7.5$, in which case none of our
discs should or was found to fragment.

\subsection{Initial conditions}

We consider a system comprising a central star, modelled as a point
mass with mass $M_{\star}$, surrounded by a disc with mass
$M_{\mathrm{disc}}$. We have performed three simulations with mass
ratios $q=M_{\mathrm{disc}}/M_{\star}$, of $0.05$, $0.1$, and $0.25$,
respectively. The initial surface density profile is taken to be a
power-law $\Sigma\propto R^{-1}$, while the initial temperature
profile was $T\propto R^{-1/2}$. The initial velocity profile was
calculated by including the enclosed cylindrical mass when determining
the angular frequency $\Omega$. With these initial conditions, the
minimum value of $Q$ is attained at the outer edge of the disc. For
each simulation the temperature normalisation is chosen such that the
minimum value of $Q$ is $Q_{\mathrm {min}}=2$, and the whole disc is
initially gravitationally stable.  The disc is assumed to be in
vertical hydrostatic equilibrium (see, for example,
\citealt{pringle81}), and we compute the scale height, $H$, using
$H=c_s/\Omega$ and distribute the particles such that the vertical
density profile is a Gaussian. Actually, in a self-gravitating disc,
the vertical density profile is not rigorously Gaussian \citep{BL99},
so our initial setup, strictly speaking, is not in dynamical
equilibrium. However, dynamical equilibrium is achieved rapidly (i.e.,
in a dynamical timescale) during the simulation.

Our calculations are essentially scale-free.  In dimensionless units,
the disc extends from $R_{\mathrm{in}}=0.25$ to $R_{\mathrm{out}}=25$,
and we have taken $M_{\star}=1$.  In these units, one dynamical
timescale (orbital period) at $R=1$ is equal to $2\pi$ code
units. Therefore, one orbital period at the outer edge of the disc is
roughly equal to 800 time units.

The disc is not initially in thermal equilibrium since the main source
of heating, i.e. gravitational instabilities, is turned off because
the disc is initially stable, while cooling is already effective. We
follow our simulations for $\approx 5000$ time units
(i.e. approximately $6$ orbital periods at the outer edge of the
disc), by which stage the whole disc has reached thermal equilibrium.

\subsection{Artificial viscosity: the Balsara switch}

\label{sec:balsara}

The standard SPH viscosity (eg., \citealt{monaghan92}) consists of a
quadratic term similar to a Von Neumann-Richtmeyer viscosity
(characterised by a dimensionless coefficient called
$\beta_{\mathrm{SPH}}$), and a linear term that introduces a viscosity
in shearing flows (characterised by a dimensionless coefficient
called $\alpha_{\mathrm{SPH}}$). Since the goal of this work is to
study transport associated with gravitational instabilities, we would
like to reduce any angular momentum transport associated purely with
the artificial viscosity (i.e., we wish to reduce the artificial shear
viscosity). 

\citet{balsara95} suggested adding a shear correction term, known as
the Balsara switch, to the standard SPH artificial viscosity which
maintains the viscosity in compressional flows but reduces it to zero
in pure shear flows. For all the simulations presented here we have
used the Balsara form of the artificial viscosity and have used a
value of 0.1 for the coefficient of the linear artificial viscosity
term ($\alpha_{\mathrm{SPH}}$) and a value of 0.2 for the coefficient
of the quadratic viscosity term. A detailed discussion of the
transport of angular momentum induced by the chosen artificial
viscosity is presented in the Appendix. Here we anticipate that the
angular momentum transport induced by artificial viscosity is at least
a factor 10 smaller than that induced by gravitational instabilities
in our simulated discs, and therefore plays a minor role in the
dynamics of the disc. 

\section{Results}
\label{sec:results}

\subsection{General features}

\begin{figure*}
\centerline{ \psfig{figure=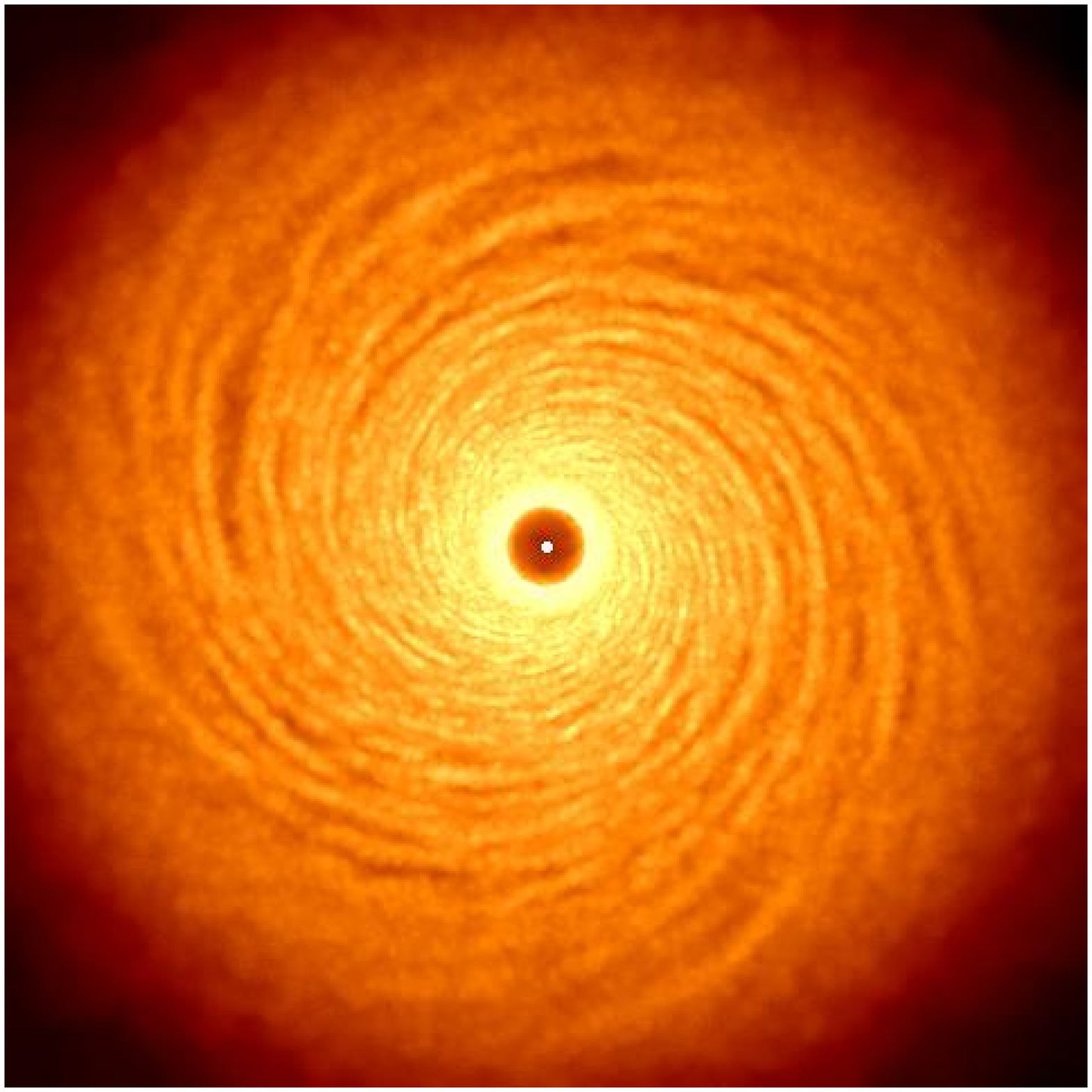,width=.4\textwidth}
             \psfig{figure=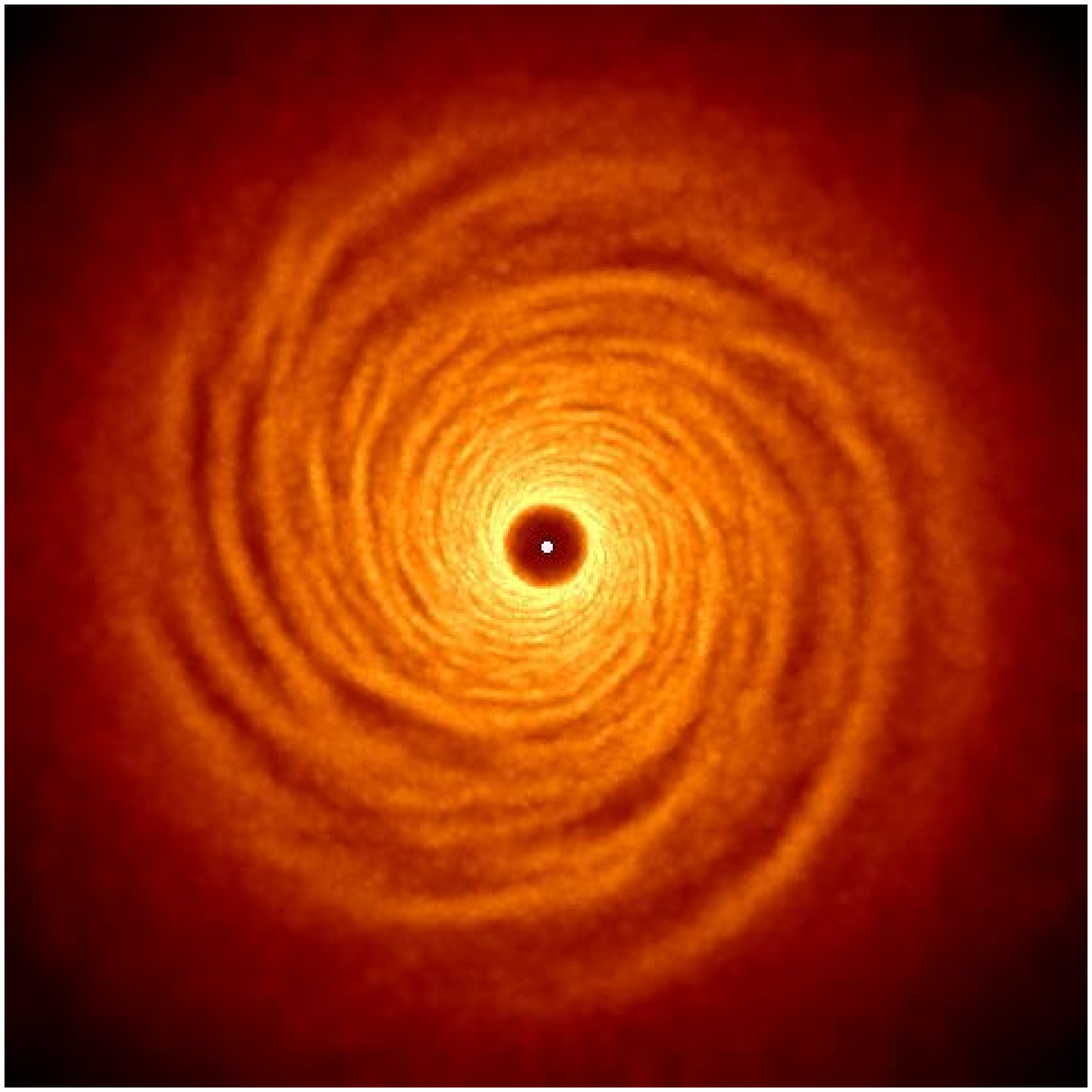,width=.4\textwidth}}
\centerline{ \psfig{figure=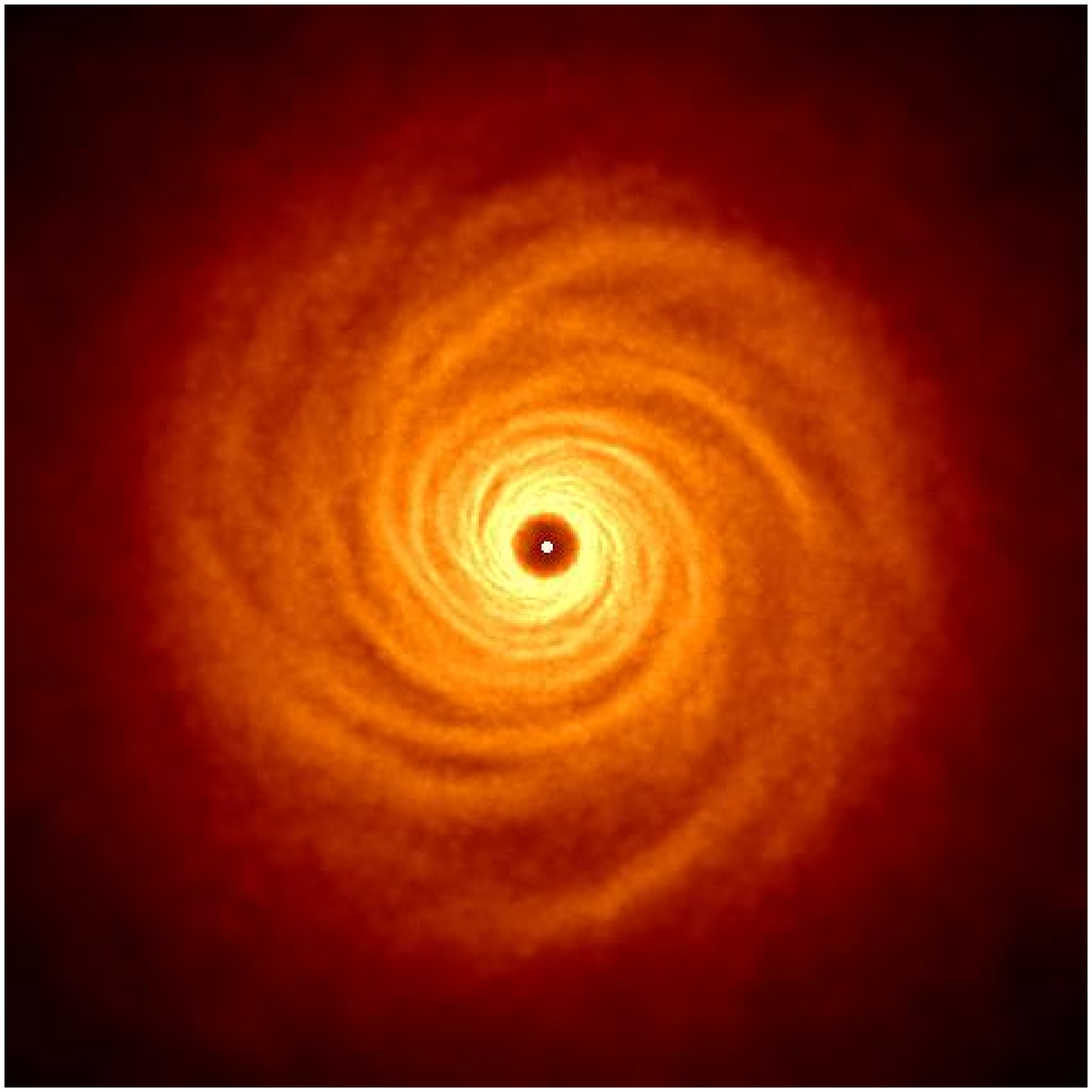,width=.4\textwidth}}
\caption{Surface density structure at the end of the simulations for
  (upper left) $q=0.05$, (upper right) $q=0.1$, and (bottom)
  $q=0.25$.}
\label{fig:density}
\end{figure*}

\begin{figure*}
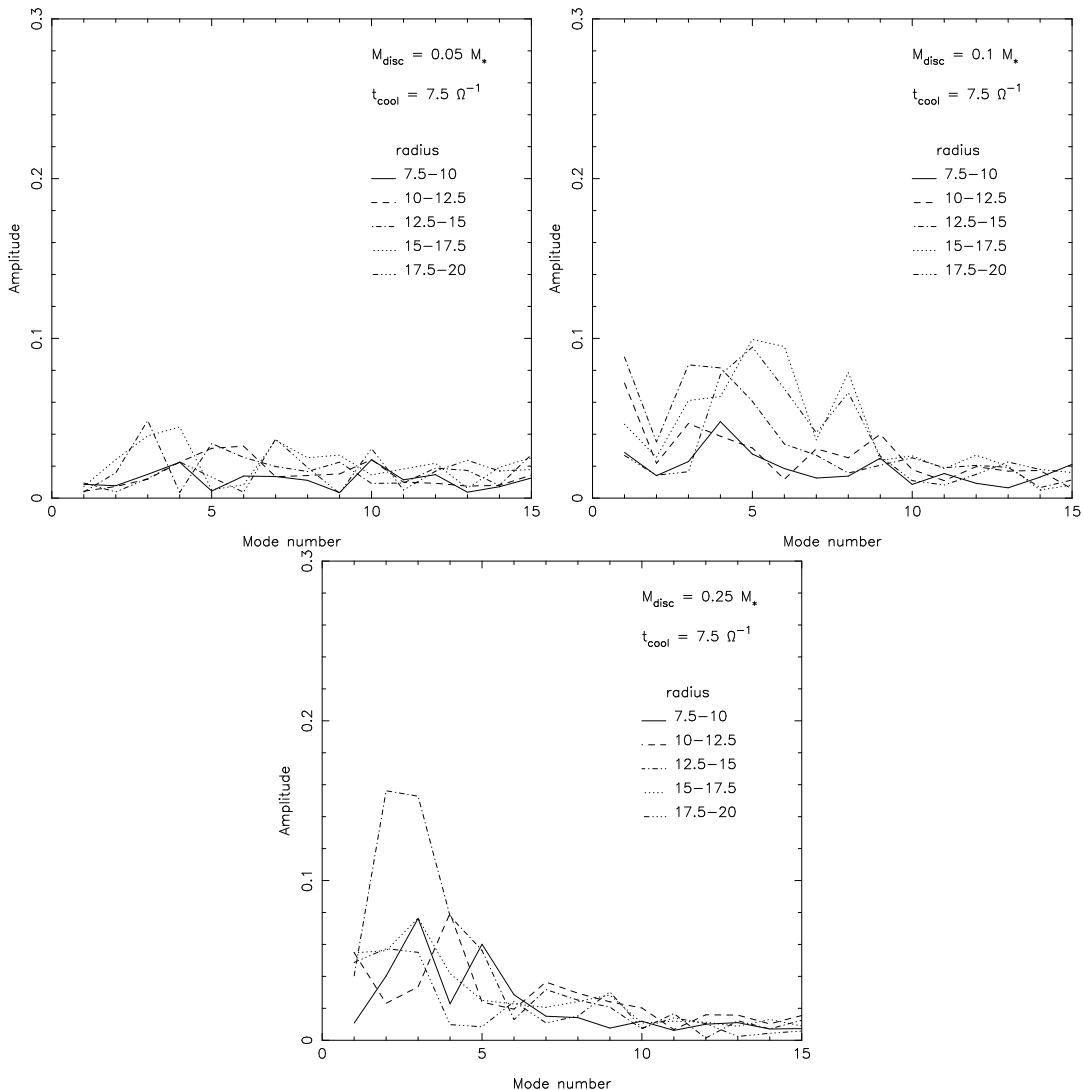

\centerline{ \psfig{figure=modeampMd005.ps,width=.4\textwidth}
             \psfig{figure=modeampMd01.ps,width=.4\textwidth}}
\centerline{ \psfig{figure=modeampMd025.ps,width=.4\textwidth}}
\caption{Amplitude  of the  first  Fourier components  of the  density
  structure for different radial ranges  in the three cases: (upper left)
  $q=0.05$, (upper right) $q=0.1$, (bottom) $q=0.25$.}
\label{fig:modes}
\end{figure*}

\begin{figure*}
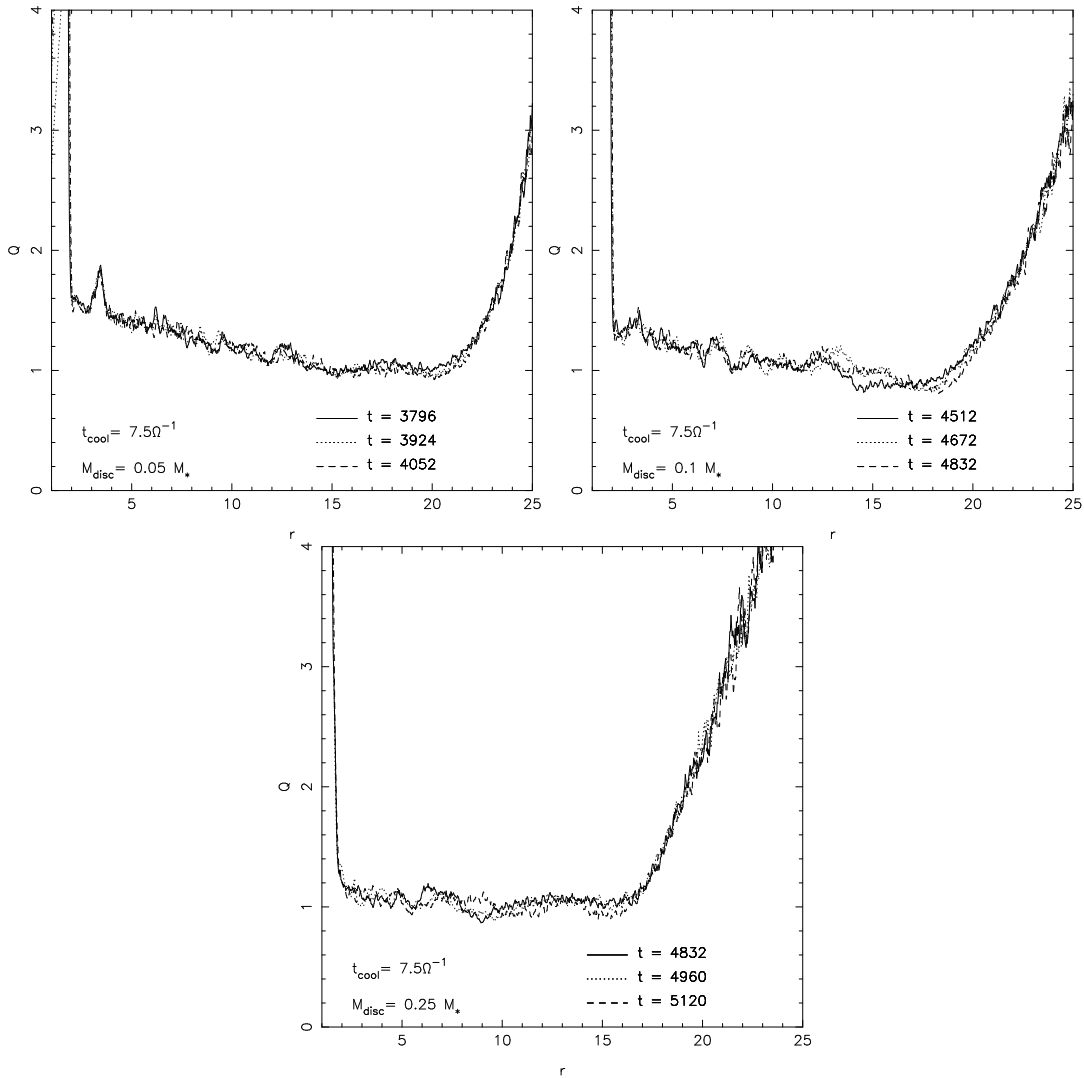

\centerline{ \psfig{figure=Md005Q_2.ps,width=.4\textwidth}
             \psfig{figure=Md01Q_2.ps,width=.4\textwidth}}
\centerline{ \psfig{figure=Md025Q_2.ps,width=.4\textwidth}}
\caption{Profiles of the $Q$ parameter for the three simulations:
  (upper left) $q=0.05$, (upper right) $q=0.1$, (bottom) $q=0.25$.}
\label{fig:Q}
\end{figure*}

\begin{figure*}
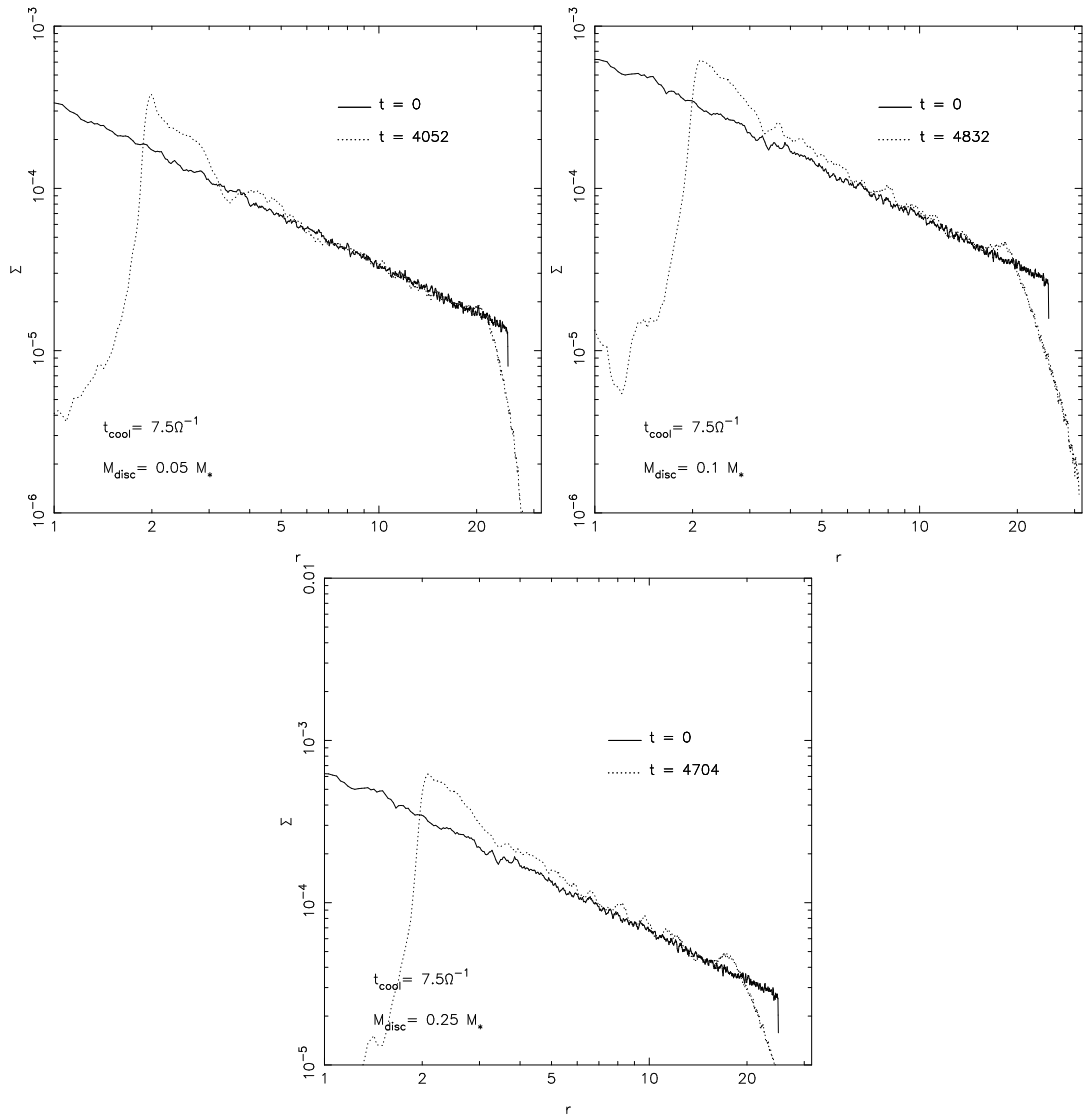

\centerline{ \psfig{figure=surfdensMd005_2.ps,width=.4\textwidth}
             \psfig{figure=surfdensMd01_2.ps,width=.4\textwidth}}
\centerline{ \psfig{figure=surfdensMd025_2.ps,width=.4\textwidth}}
\caption{Radial profiles of the surface density at the end of the
three simulations: (upper left) $q=0.05$, (upper right) $q=0.1$,
(bottom) $q=0.25$. The solid line shows the initial surface density
profile.}
\label{fig:sigma}
\end{figure*}

In all our simulations the disc initially cools down until the value
of $Q$ becomes small enough for gravitational instabilities to become
effective and to provide a source of effective heating.  At later stages
the disc settles into a quasi-steady state, where the heating provided
by the instabilities balances the imposed cooling term.  As predicted by the
argument of self-regulation, this quasi-steady state is characterised
by an almost constant value of $Q\approx 1$, over a wide radial range.
The main features of the simulations are summarised in Figs.
\ref{fig:density}, \ref{fig:modes}, \ref{fig:Q}, and \ref{fig:sigma}.

Fig.  \ref{fig:density} shows the surface density structure for the
three simulations, with different total disc mass ($q=0.05$, $q=0.1$
and $q=0.25$).  The global structure induced by self-gravity is
clearly seen in all three cases. The density enhancement in the spiral
arms $\Delta\rho/\rho$ typically ranges between 2 and 4 in all
cases. Already from this figure it can be noticed that, as the total
disc mass increases, the pattern of the instability becomes
progressively more dominated by low-$m$ modes.  This is confirmed by
the Fourier analysis shown in Fig. \ref{fig:modes}, that shows how in
the higher total mass cases the modes with $m<5$ dominate the
structure. The Fourier amplitudes in Fig. \ref{fig:modes} are computed
as follows: we divided the disc in concentric rings of width $\Delta
R=2.5$ in dimensionless units. For each ring we then compute the mode
amplitude $A_m$ as:
\begin{equation}
A_m=\left|\sum_{i=1}^{N_{\mathrm{ring}}}\frac{\mbox{e}^{-im\phi_i}}
{N_{\mathrm{ring}}}\right|,
\end{equation}
where $N_{\mathrm{ring}}$ is the number of particles in the ring, and
$\phi_i$ is the azimuthal angle of the $i$-th particle. 

Fig. \ref{fig:Q} shows the radial profile of $Q$ at three different
times, towards the end of the simulation.  The profile is not
significantly altered with time, indicating that the simulations have
reached thermal equilibrium, and are in a quasi-steady
state. Actually, the fact that the self-regulated value of $Q$ turns
out to be so close to unity in our simulations is quite remarkable.
In fact, the marginal stability value $\bar{Q}$ is equal to unity only
when the perturbation analysis is restricted to the case of the {\it
axisymmetric} stability of an {\it infinitesimally thin} disc.  The
disc thickness has an important stabilising effect, which leads to a
lower temperature (and thus a lower value of $Q$) of the marginally
stable state; on the other hand, even a light disc can be destabilised
by the presence of swing amplified non-axisymmetric disturbances with
high $m$ (for a discussion of these mechanisms see, e.g.,
\citealt{bertinbook2}).  It then appears that these two competing
effects basically counterbalance each other.  A detailed investigation
of the combined effect of these physical mechanisms in determining the
precise value of $Q$ at marginal stability is, however, beyond the
scope of the present paper.

Fig. \ref{fig:sigma} displays the initial and final radial profiles of
the surface density (averaged in the azimuthal direction). It can be
noticed that there is no significant evolution of the surface density
at large radii. The major changes in the surface density occur close to
the boundaries. In particular, there is a rapid drop in surface
density close to the inner boundary. This is due to the fact that we
have not attempted to give a detailed description of the boundary
layer. The sudden lack of SPH particles at $R<0.25$ causes an
increased artificial pressure which pushes the inner particles into
the sink radius of the star. This effect should be reduced with a more
accurate description of the boundary layer. We have also checked that
the total angular momentum of the disc is conserved to within a few
percent throughout all the simulations.

When low values of the artificial viscosity are used, particle
interpenetration might lead to a poor representation of strong shocks
in SPH. This is not a serious issue in our case, since in our
simulations only mildly supersonic shocks are involved. Based on the
density contrast in the spiral arms, we estimate the Mach number of
the shocks to be $\mathcal{M}\lesssim 1.5$. At these low values of
$\mathcal{M}$, a value of $\beta_{\mathrm{SPH}}\approx 0.2$ is already
sufficient to stop particle interpenetration \citep{batephd}. This is
confirmed by the well defined spiral structure that we obtain (that
would have been smeared out if significant particle interpenetration
was indeed present), consistent with the results of previous
simulations that used the standard SPH viscosity and higher values for
the viscosity coefficients \citep{rice03a,rice03b}.

\begin{figure*}
\centerline{ \epsfig{figure=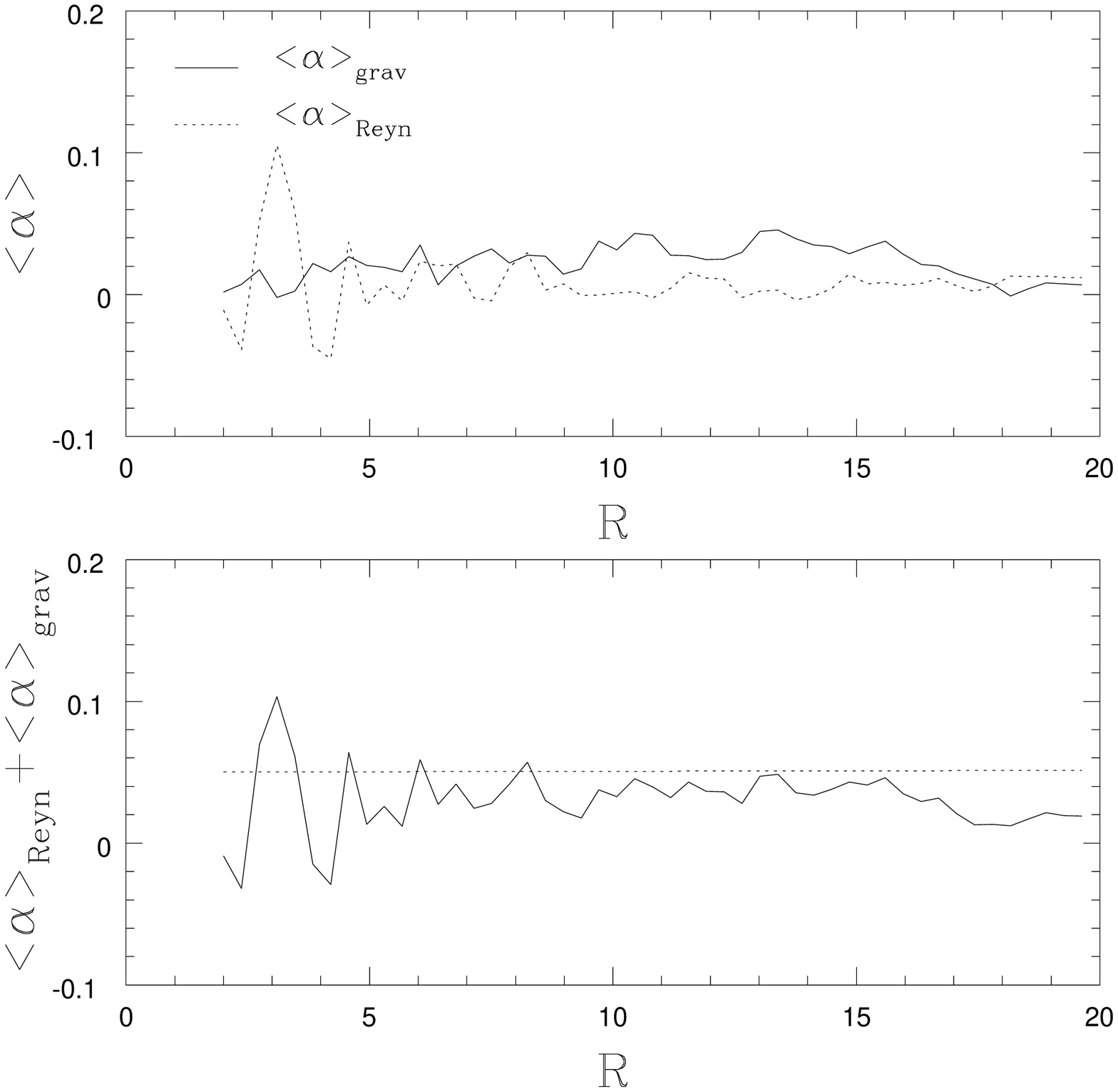,width=.4\textwidth}
             \epsfig{figure=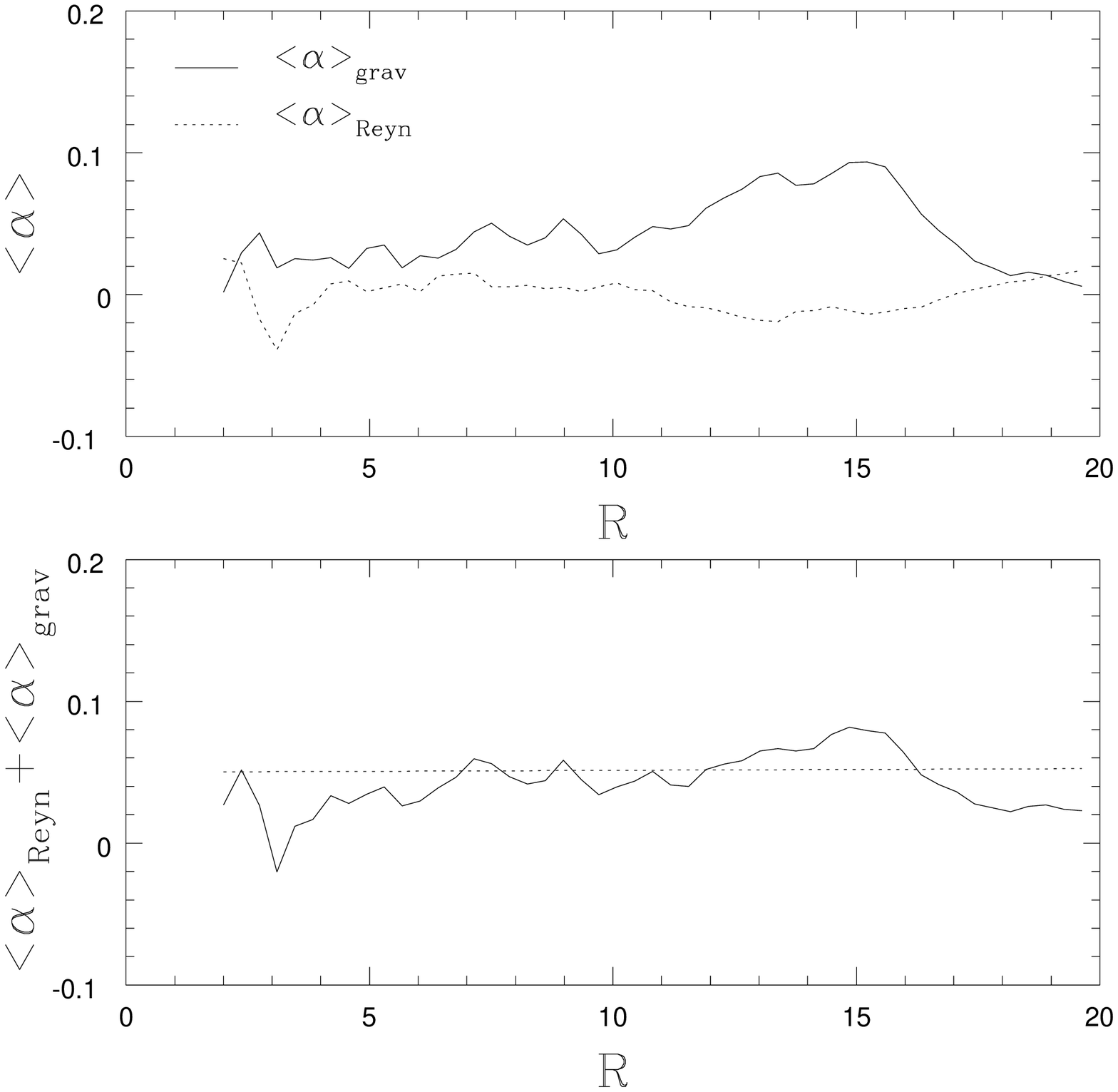,width=.4\textwidth}}
\centerline{ \epsfig{figure=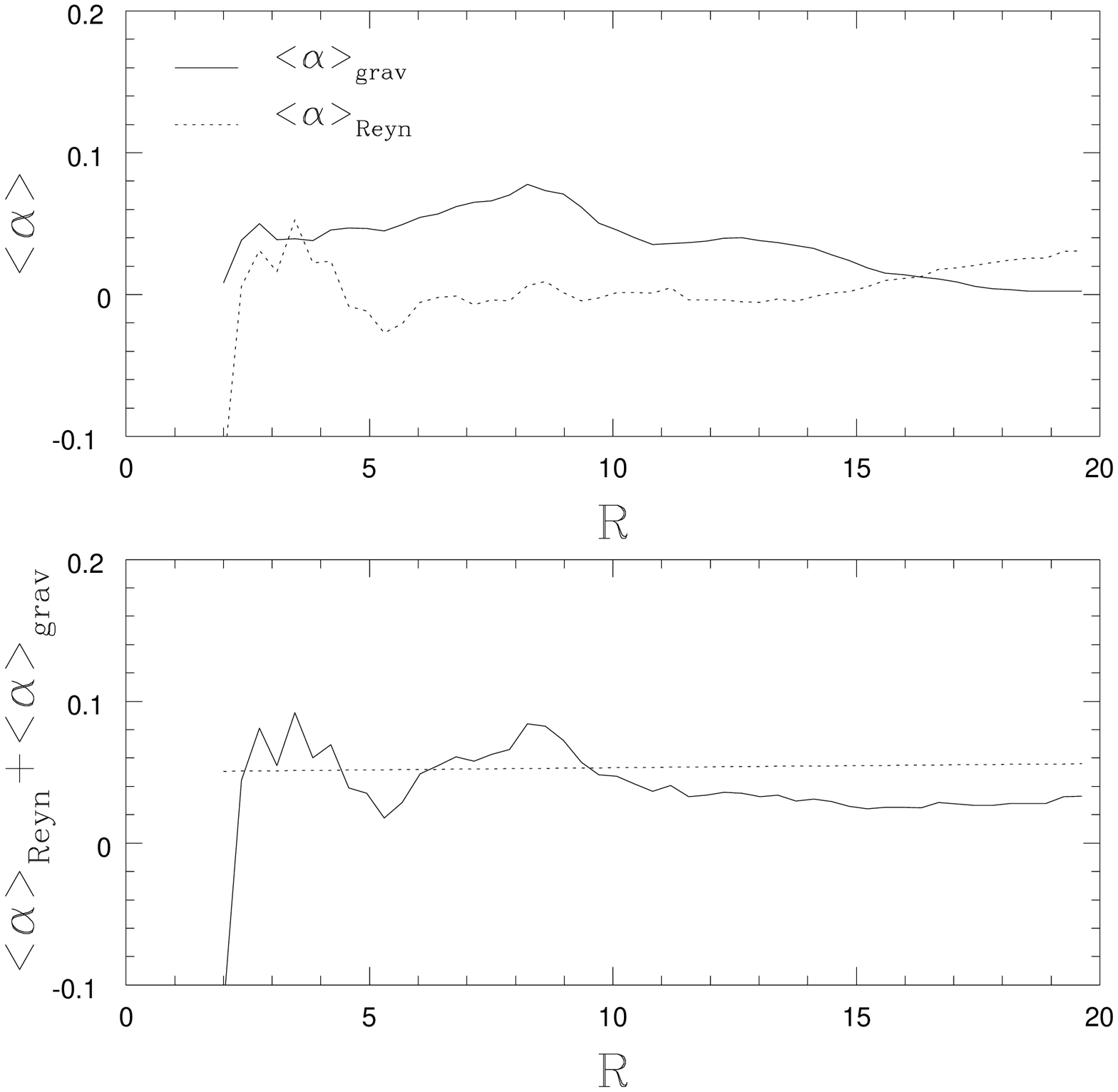,width=.4\textwidth}}
\caption{Effective $\alpha$ produced by gravitational instabilities
 for (upper left) $q=0.05$, (upper right) $q=0.1$, and (bottom)
 $q=0.25$. The top panel shows the separate contribution of
 $\alpha_{\mathrm {grav}}$ and $\alpha_{\mathrm{Reyn}}$, the lower
 panel shows the sum of the two contributions compared with the
 expected value from a local viscous model (dotted line).}
\label{fig:alpha}
\end{figure*}

\begin{figure*}
\centerline{ \epsfig{figure=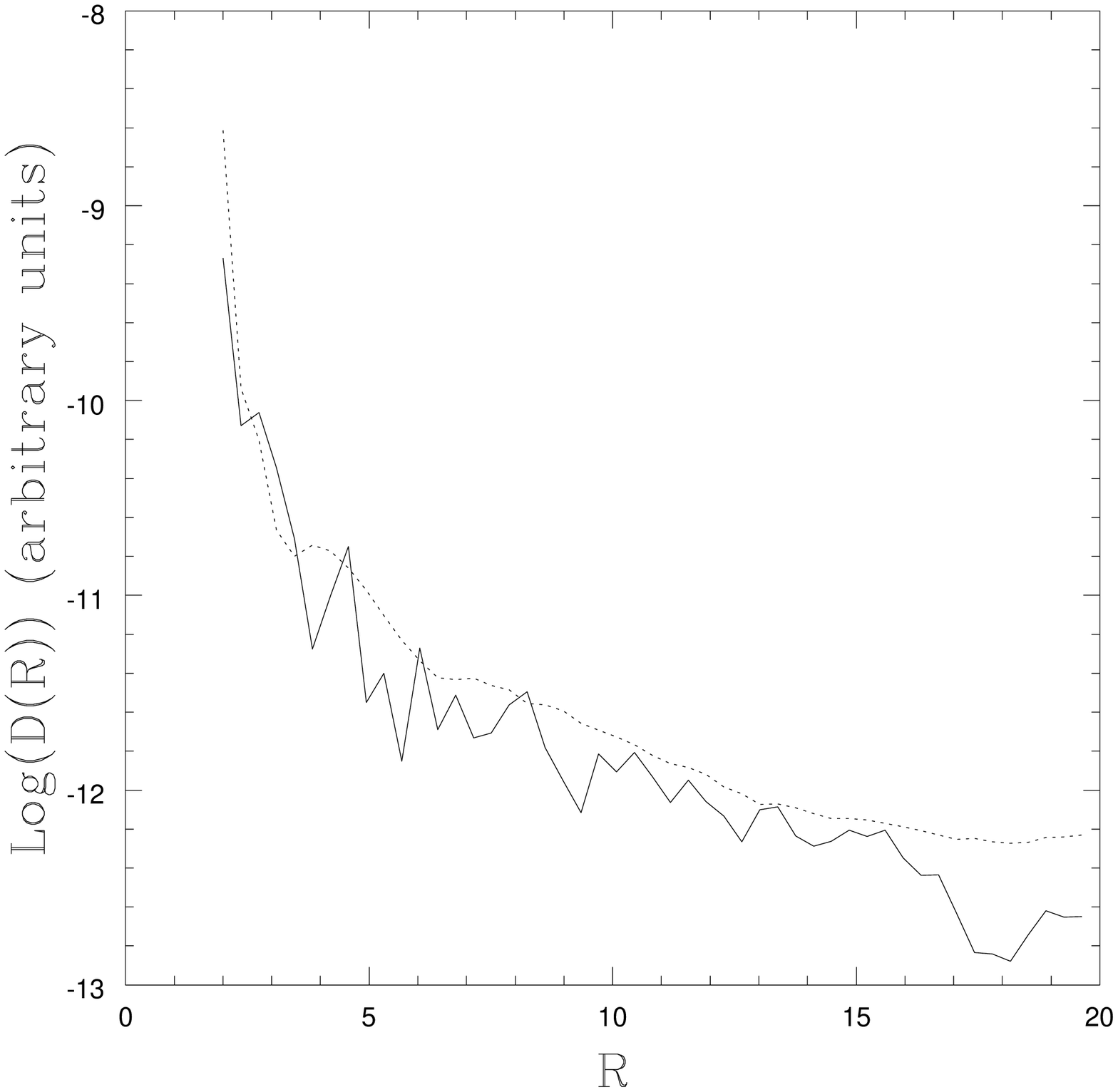,width=.4\textwidth}
             \epsfig{figure=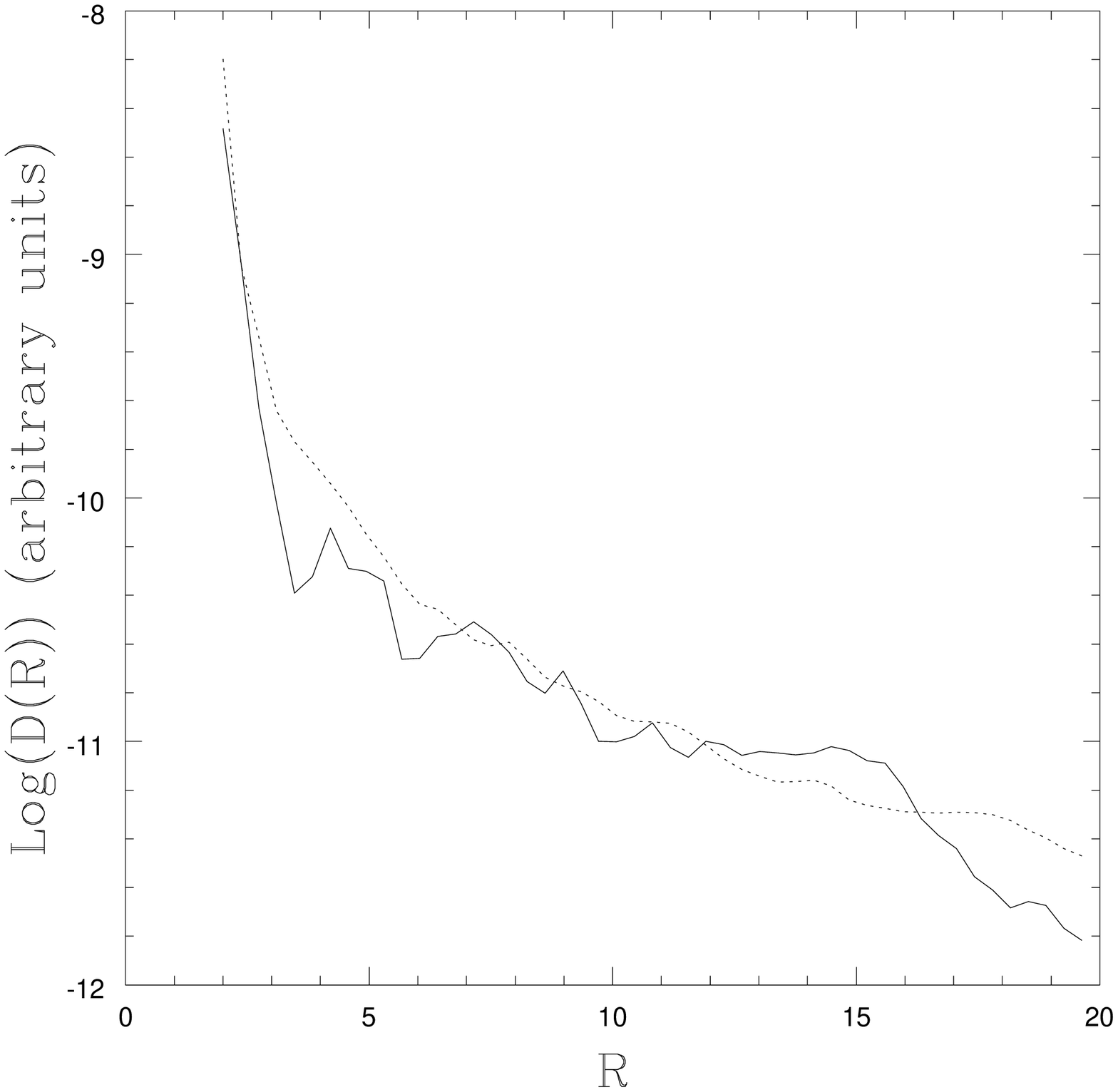,width=.4\textwidth}}
\centerline{ \epsfig{figure=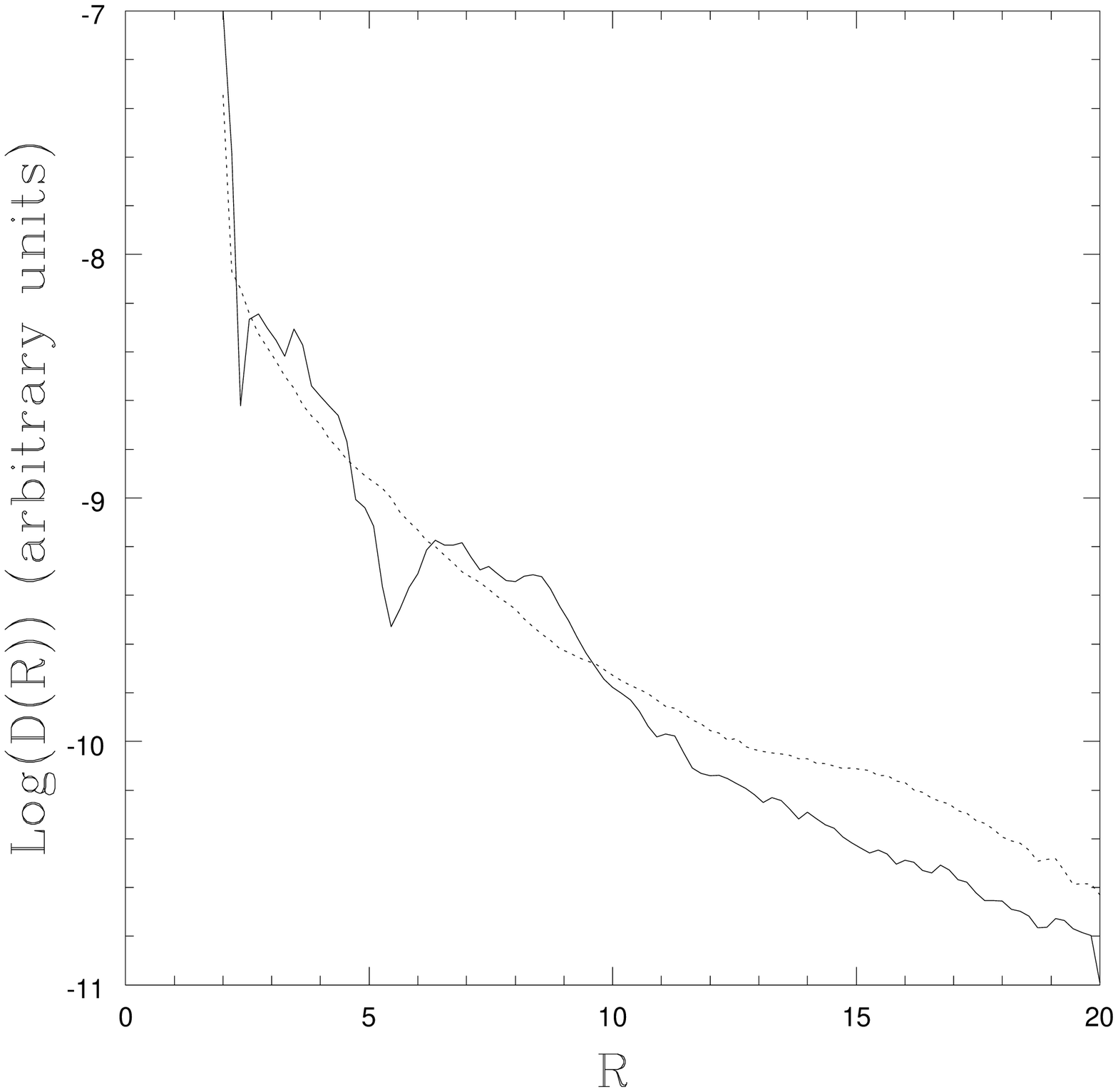,width=.4\textwidth}}
\caption{Viscous dissipation rate (solid line, from
 Eq. (\ref{dissipation}) and actual cooling rate (dotted line) for
 (upper left) $q=0.05$, (upper right) $q=0.1$, and (bottom) $q=0.25$.}
\label{fig:dissipation}
\end{figure*}

\subsection{Angular momentum transport and energy dissipation}

The torque produced by gravitational instabilities in the disc is
given by the sum of the two terms described in Eqs. (\ref{stress}) and
(\ref{hydro}).  After averaging the stress tensor azimuthally and
radially, over a small region $\Delta R=0.1$, we compute the
corresponding value of $\alpha$ (see Eq. (\ref{alpha})):
\begin{equation}
\alpha(R)=\left|\frac{\mbox{d}\ln\Omega}{\mbox{d}\ln
R}\right|^{-1}\frac{<T_{R\phi}^{\mathrm{grav}}>+<T_{R\phi}^{\mathrm{Reyn}}>}
{\Sigma c_s^2}.
\label{alpha2}
\end{equation} 

The resulting radial profiles of $\alpha$ are shown in Fig.
\ref{fig:alpha} for the three cases $q=0.05$, $q=0.1$, and $q=0.25$.
The upper panels show separately the hydrodynamic and gravitational
contributions to $\alpha$, while the bottom panels show the sum of the
two. The plots show the time average of $\alpha$ at the end of the
simulation, once the disc has reached a quasi-steady state. The
time-averaging interval is 500 time units, i.e. 0.6 orbital times at
the outer disc edge. 

We can now use the general results of viscous disc theory outlined in
Sec.  \ref{sec:general} to test the locality of transport in our
simulations. In fact, Eq.  (\ref{alphavisc}) gives us firm
expectations for the value of $\alpha$ needed to balance the imposed
cooling, {\it if energy dissipation can be treated in a viscous
framework}, i.e.  by using Eq.  (\ref{dissipation}).  In particular,
in our simulations $t_{\mathrm{cool}}\Omega=\beta=7.5$, $\gamma=5/3$
and, since our discs are nearly Keplerian, $\de\ln\Omega/\de\ln
R\approx -3/2$.  Inserting these numbers in Eq. (\ref{alphavisc})
would give us an expected value of $\alpha\approx 0.05$.  It is
important to note that the fact that the expected $\alpha$ turns out
to be nearly independent on radius is a result of choosing the cooling
time to be simply proportional to the dynamical time-scale.  In the
general case, of course, the resulting $\alpha$ needs not be
constant. The dotted line in Fig. \ref{fig:alpha} shows the expected
value of $\alpha$.  Our results are in fairly good agreement with the
expectations of viscous transport theory.

It is also interesting to compare the dissipation rate $D(R)$ that
would result if the transport process were viscous, i.e. computing
$D(R)$ based on Eq. (\ref{dissipation}), with the actual dissipation
rate. Since our discs are in thermal equilibrium, the dissipation rate
can be obtained from the cooling rate using Eq. (\ref{cooling}). This
comparison is shown in Fig. \ref{fig:dissipation}, where the solid
line shows $D(R)$ computed from Eq. (\ref{dissipation}), while the
dotted line shows the cooling rate. Again, the plots show a good
agreement with the expectations, the only exception being the outer
parts of the most massive disc, where the actual dissipation rate
seems to be slightly larger than that expected from a viscous process.

In general, our results indicate that, up to disc masses of
$M_{\mathrm{disc}}=0.25M_{\star}$, gravitationally induced transport
is reasonably well described within a local framework.  This result
could also be anticipated since the disc dynamics in all the three
simulations are dominated by rather high-$m$ modes, that dissipate on
a short length-scale.

As a separate test for the locality of the transport, we have also
computed $\alpha_{\mathrm{part}}(R,d)$, which we define as the
gravitational part of $\alpha(R)$, taking into account only those
particles inside a spherical radius $d$ from the radial point $R$
where we compute the stress. This quantity gives us a measure of the
size of the region that has the most significant contribution to the
gravitational stress at a given point. However, it should be kept in
mind that this quantity only accounts for the stress produced by the
gravitational field, without including the hydrodynamical component.

\begin{figure}
\centerline{ \epsfig{figure=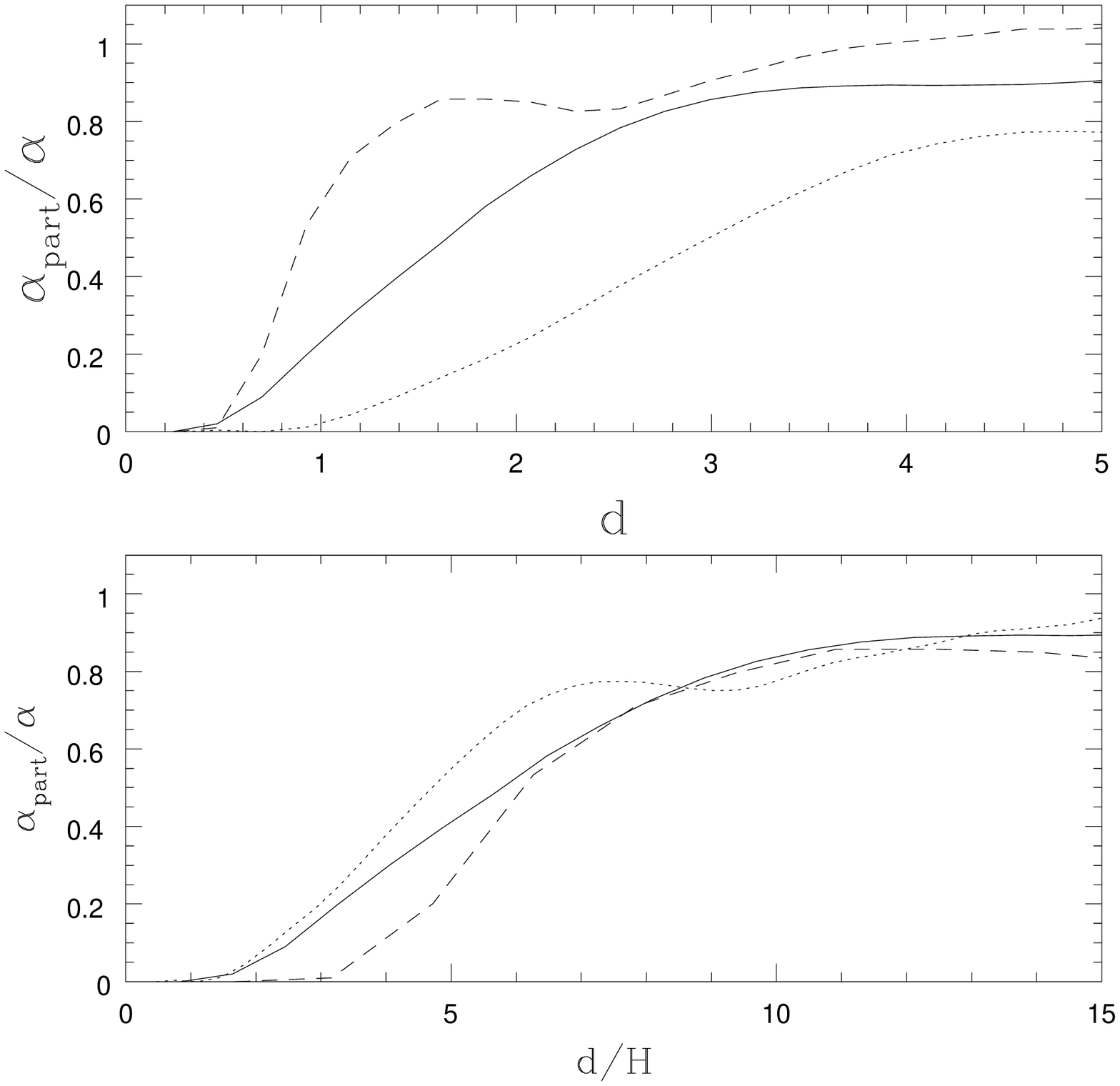,width=84mm}}
 \caption{Contribution to the stress  at $R=15$  from regions  of the
 discs at a distance $<d$  from $R$. Dashed line: $q=0.05$, solid line:
 $q=0.1$. Dotted line: $q=0.25$}
\label{fig:part}
\end{figure}

Fig.  \ref{fig:part} shows the results for $R=15$ for the three
simulations. The upper panel shows clearly that the more massive the
disc, the more distant regions from the point contribute to the
stress. However, it should be noted that, since in the thermal
equilibrium state $Q\approx 1$ in all the three cases, a more massive
disc is also characterised by a larger value of the effective thermal
speed $c_s$, and is therefore thicker. Since we expect that
gravitational disturbances propagate over a length-scale of the order
of the thickness of the disc, to compare appropriately the results of
the three simulations, we should check the dependence of
$\alpha_{\mathrm{part}}$ on $d/H$. This is shown in the bottom panel
of Fig. \ref{fig:part}. In all cases, more than 80\% of the
contribution to the gravitational torque comes from a region with size
$\Delta\approx 10H$. Transport is local if $\Delta\ll R$.  We can
therefore conclude that angular momentum transport is determined by
local conditions only for discs with $H/R \ll 0.1$. When $Q\approx 1$,
the ratio $H/R$ is proportional to $M_{\mathrm{disc}}/M_{\star}$,
which means that sufficiently massive discs may violate the previous
condition. Actually, our most massive disc (for which
$M_{\mathrm{disc}}/M_{\star}=0.25$) only marginally satisfies $H/R<
0.1$. Indeed, as can be seen from Figs. \ref{fig:alpha} and
\ref{fig:dissipation}, global wave transport might play some role in
the outer disc, in this case.

\section{Discussion}
\label{sec:discussion}

\subsection{Comparison with previous work}

The distinctive feature of the present work is that we have performed
global, three-dimensional simulations of massive discs, including the
detailed effects of heating from gravitational instabilities and
cooling. Our simulations are similar to those performed by
\citet{rice03a} and \citet{rice03b}, with the difference that these
previous investigations were concerned about either the motion of the
central object caused by the massive disc, or the issue of
fragmentation of the disc, while here the main goal is to characterise
the transport properties induced by gravitational instabilities.

We believe that both performing 3D simulations and solving explicitly
the energy equation are essential to determine the final outcome of
the instabilities. In fact, the dynamical properties of
self-gravitating discs are determined to a large extent by the process
of self-regulation, which is strongly dependent on the detailed
heating and cooling mechanisms, as outlined in the
Introduction. Furthermore, since one of the main tests we want to
perform is to check the kind of dissipative process associated with
gravitational instabilities, solving the energy equation is
essential. The requirement of 3D simulation is also fundamental, since
the typical size of gravitational disturbances is related to the disc
thickness, so that any zero-thickness simulation will automatically
lead to an underestimate of global effects.

Previous numerical work done on the subject includes global, 3D SPH
simulations of massive {\it isothermal} discs \citep{laughlin94};
global, 2D, SPH simulations with detailed heating and cooling
\citep{nelson2000}; global, 3D, grid-based simulations with heating
and cooling \citep{pickett2000}; and local, 2D grid-based simulations
with heating and cooling \citep{gammie01}. In this Section we describe
the similarities and the differences between our study and these
previous ones.

One of the first studies of gravitational instabilities in discs was
performed by \citet{laughlin94}. They modelled a very massive disc
(with $M_{\mathrm{disc}}\approx M_{\star}$) and followed its evolution
with a 3D SPH code, without including any heating or cooling term, but
simply assuming that the disc was locally isothermal. In this study,
the authors also tried to give a detailed characterisation of the
transport. Their approach was however slightly different to ours, in
that they evolved their simulation long enough to capture the viscous
evolution of the disc, and then compared the evolution of the
azimuthally averaged surface density with simple one-dimensional
viscous models, concluding that the disc evolution could be well
reproduced by a viscous model with $\alpha\approx 0.03$. This work is
important because it clarifies that gravitational instabilities are
actually able to transport angular momentum efficiently and that the
surface density evolution is indeed of a diffusive nature, as expected
(see Section \ref{sec:general}), but does not answer the important
question of whether energy dissipation is local or global.

\citet{nelson2000} performed 2D simulations with particular emphasis
on the cooling processes in the disc and included a more realistic
cooling function than the simple parameterisation adopted here. Their
disc mass was $M_{\mathrm{disc}}=0.2M_{\star}$, very similar to our
most massive case. They estimated the effective $\alpha$ associated
with artificial viscosity (see Section \ref{sec:balsara}) to be of the
order of $5\times 10^{-3}$, comparable to the expectations based on
Eqs. (\ref{eq:visc}) and (\ref{eq:alphaartificial}), and an order of
magnitude smaller than what we get here from transport induced by
gravitational torques. Indeed, in our simulations most of the angular
momentum is carried by collective instabilities and by gravitational
torques, rather than by the artificial shear. Comparing the
dissipation rates from shock heating with that from turbulent heating
in their simulations, \citet{nelson2000} conclude that, at least in
the outer disc, gravitational torques should not contribute more than
the torques associated with artificial viscosity, while in our
simulation gravitational torques are everywhere dominant. A possible
reason for this discrepancy lies in the 3D nature of our simulations,
which may allow gravitational disturbances to travel further and
allows low-$m$ modes to be more prominent. Actually, even if the mode
amplitudes that we obtain here (see Fig. \ref{fig:modes}) are in rough
agreement with what obtained by \citet{nelson2000}, they find very
similar mode amplitudes for all modes with $m<8$, while in our case
there is a marked increase of mode amplitude for modes with $m<5$ (see
Fig. \ref{fig:modes}).

\citet{gammie01} performed an analysis very similar to our own
(actually, we adopt the same prescription for the cooling term, see
Eq. (\ref{cooling2})). The main difference is that Gammie's
simulations are local and two-dimensional while ours are global and
three-dimensional. He computes the effective $\alpha$ based on
Eq. (\ref{alpha2}), as we do. The main results that we obtain are
basically in agreement with Gammie's results, in that the expected
value of $\alpha$ from a viscous theory of discs is very close to the
computed value in the simulations. However, by using a global
approach, we are now able to check how these results depend on the
total disc mass and on the disc thickness, whereas Gammie could only
extrapolate his results to thicker configurations.

Recently, \citet{gammie03} have extended the work of \citet{gammie01}
to include a more realistic cooling function in 2D simulations.
Actually, the use of detailed cooling functions makes it even more
important to perform a full 3D simulation, since most of the radiative
transport will occur in the vertical direction, so that using
vertically averaged values for the relevant physical quantities, as
done by \citet{gammie03}, might lead to incorrect results.

\subsection{Impact on observed systems}

It is now commonly accepted that most low-mass young stellar objects
possess a circumstellar disc, with lifetimes of at least
1Myr. Circumstellar discs play an important role in the process of
star formation and are believed to be the site where planet formation
occurs. Observational evidence for the presence of such discs is
either indirect, i.e. based on the disc emission at long-wavelengths,
such as in the infrared (starting from \citealt{adams88}) and at
sub-mm wavelengths \citep{beckwith90}, or direct, i.e. by imaging of
{\it silhouette} discs in the Orion nebula \citep{mccaughrean96}.
Especially in the earliest phases of star formation these discs might
be fairly massive: for example, \citet{launhardt2001}, using sub-mm
observations, report the discovery of a massive disc (with
$M_{\mathrm{disc}}/M_{\star}\gtrsim 0.3$) in a very young (Class I)
protostellar object. There are also some indications that the discs in
FU Orionis objects might be fairly massive: \citet{sandell2001} report
$M_{\mathrm{disc}}/M_{\star}\gtrsim 0.1$ in most of the FU Orionis
discs they have observed. As already mentioned, detailed modelling of
FU Orionis outbursts produce radial profiles of $Q$ that fall below
unity already at a distance of $\approx 1$ au from the central object
\citep{bellin94}. All these systems are likely to be affected by
self-gravity that, if indeed energy dissipation is non-local, may
produce some observable modification in the SED. It is then
interesting to see that the models by \citet{bellin94} predict
$H/R\gtrsim 0.1$, which, according to the results of this work, are
large enough for non-local effects to become important, as suggested
by \citet{LB2001}.

In the context of AGN discs, the distance at which the disc becomes
marginally stable to gravitational instabilities is typically of the
order of $10^3R_{\mathrm{g}}$ \citep{LB03a}, where $R_{\mathrm{g}}$ is
the gravitational radius of the black-hole. Water maser emission
\citep{greenhill97} and radio continuum observations
\citep{gallimore97} show that in many cases the disc can extend to
radii much larger than that, thus allowing self-gravity to influence
the disc structure. In this context, self-regulated models have been
applied both to the modelling of the SED \citep{sirko03} and to the
modelling of the rotation curve in the outer disc of the Seyfert galaxy
NGC 1068 \citep{LB03a}. \citet{gammie03} use their 2D results with
``realistic'' cooling (see previous Subsection) to argue that the disc
model proposed by \citet{LB03a} for NGC 1068 would be subject to
fragmentation, but unfortunately \citet{gammie03} do not explore the
region of the parameter space relevant to the Lodato \& Bertin model.

As a final comment, we note that the present work refers to the
situation where the dominant source of transport and dissipation is
provided by gravitational instabilities. In observed systems, other
sources of transport could be present, which might reduce the effect
of gravitational instabilities.

\section{Conclusions}
\label{sec:conclusions}

In this paper we have investigated the transport properties induced by
disc self-gravity in relatively massive accretion discs. In
particular, we have discussed the extent to which angular momentum
transport and energy dissipation can be described within a viscous
framework. To this goal, we have performed global, three-dimensional
simulations using SPH.

Most of the recent work on the subject has focused on the issue of the
conditions for disc fragmentation, especially in relation to the
process of planet formation in protostellar discs
\citep{boss02,rice03b,mayer02}, or massive star formation in AGN discs
\citep{goodman03,levin03}. Here, we would like to check the behaviour
of self-gravitating discs in the case where fragmentation does not
occur and the disc evolves towards a quasi-steady state where cooling
is balanced by heating from gravitational instabilities.

The issue of the transport properties induced by gravitational
instabilities has been discussed analytically in the past by
\citet{lyndenbell72} and, more recently, by \citet{balbus99}. There
are two major aspects of the problem: {\it (i)} to what extent is the
shear stress $T_{\mathrm{R\phi}}$ at a given radius $R$ determined by
local conditions?  This point directly calls into question the use of
the $\alpha$ formalism, which explicitly requires that
$T_{\mathrm{R\phi}}$ is only dependent on the local values of density
and thermal velocity (see also the discussion in Section
\ref{sec:self}). The second aspect is {\it (ii)} whether energy
transport is only determined by the gravitationally induced shear
stress or whether instead global wave transport occurs, hence
influencing energy dissipation in the disc, as argued by
\citet{balbus99}.

We have performed several simulations, with different values of the
ratio $M_{\mathrm{disc}}/M_{\star}$, to test how the global transport
properties of the disc change with increasing total disc mass. Our
simulations appear to evolve to a quasi-steady state, characterised by
an almost flat profile of the axisymmetric stability parameter
$Q\approx 1$, in which the heating provided by gravitational
instabilities balances the imposed cooling rate. We have computed the
torque induced by gravitational instabilities and the corresponding
viscous dissipation rate, which we then compare to the actual
dissipation rate in our simulated discs. We have found values of
$\alpha\approx 0.05$, in reasonable agreement with the expectation
from a viscous theory. Energy dissipation in our simulations is also
fairly well described using a viscous approach. These results directly
address aspect {\it (ii)}, described above, and confirm the argument
of \citet{balbus99}, who had indeed argued that for ``self-regulated''
discs, in which $Q\approx 1$, global wave transport of energy would
not play a major role.

On the other hand, we have indeed noticed that more massive discs tend
to be dominated by lower $m$ modes, leading to a more global pattern
of the gravitational disturbances. We have also been able to determine
the size of the region that mostly contributes to the torque at a
given point in the disc. More than 80\% of the torque is produced
within a region of size $\Delta\approx 10H$, where $H$ is the
thickness of the disc. These results directly address issue {\it (i)},
described above. We can therefore conclude that a viscous description
of the transport in self-gravitating discs is only appropriate when
$H/R\lesssim 0.1$. In systems like FU Orionis, where models predict
disc thickness $\approx 0.1$ \citep{bellin94}, global effects could
play a role and modify the dissipation rates in the outer disc.

The results of this work should be extended both toward a more
thorough investigation of the parameter space (in particular, it is
very important to test the transport properties of discs more massive
than $0.25M_{\star}$, and to explore the results obtained with
different initial conditions and cooling timescale, which in the
present work has been taken to be $t_{\mathrm{cool}}=7.5\Omega^{-1}$),
and toward a more realistic description of the cooling function (see,
for example, \citealt{nelson2000} and \citealt{gammie03}).

\section*{acknowledgements}
The simulations presented in this work were performed using the UK
Astrophysical Fluid Facility (UKAFF). We thank C. Clarke, J.  Pringle
and G. Bertin for several interesting discussions and for a careful
reading of the manuscript. WKMR acknowledges support from a UKAFF
Fellowship. 

\bibliographystyle{mn2e}
\bibliography{lodato}

\begin{thebibliography}{}

\bibitem[\protect\citeauthoryear{Adams, Lada \& Shu}{Adams
  et~al.}{1988}]{adams88}
Adams F.~C.,  Lada C.~J.,    Shu F.~H.,  1988, ApJ, 326, 865

\bibitem[\protect\citeauthoryear{Armitage, Livio \& Pringle}{Armitage
  et~al.}{2001}]{armitage2001}
Armitage P.~J.,  Livio M.,    Pringle J.~E.,  2001, MNRAS, 324, 705

\bibitem[\protect\citeauthoryear{Artymowicz \& Lubow}{Artymowicz \&
  Lubow}{1994}]{lubow94}
Artymowicz P.,  Lubow S.~H.,  1994, ApJ, 421, 651

\bibitem[\protect\citeauthoryear{Balbus \& Hawley}{Balbus \&
  Hawley}{1998}]{balbusreview}
Balbus S.~A.,  Hawley J.~F.,  1998, Reviews of Modern Physics, 70, 1

\bibitem[\protect\citeauthoryear{Balbus \& Papaloizou}{Balbus \&
  Papaloizou}{1999}]{balbus99}
Balbus S.~A.,  Papaloizou J. C.~B.,  1999, ApJ, 521, 650

\bibitem[\protect\citeauthoryear{Balsara}{Balsara}{1995}]{balsara95}
Balsara D.~S.,  1995, Journal of Computational Physics, 121, 357

\bibitem[\protect\citeauthoryear{Bate}{Bate}{1995}]{batephd}
Bate M.,  1995, PhD thesis, University of Cambridge, UK

\bibitem[\protect\citeauthoryear{Bate, Bonnell \& Price}{Bate
  et~al.}{1995}]{bate95}
Bate M.~R.,  Bonnell I.~A.,    Price N.~M.,  1995, MNRAS, 277, 362

\bibitem[\protect\citeauthoryear{Beckwith et~al.,}{Beckwith
  et~al.}{1990}]{beckwith90}
Beckwith S. V.~W.,  et~al., 1990, AJ, 99, 924

\bibitem[\protect\citeauthoryear{Bell \& Lin}{Bell \& Lin}{1994}]{bellin94}
Bell K.~R.,  Lin D. N.~C.,  1994, ApJ, 427, 987

\bibitem[\protect\citeauthoryear{Benz}{Benz}{1990}]{benz90}
Benz W.,  1990, in Buchler J.,  ed., The Numerical Modeling of Nonlinear
  Stellar Pulsations Kluwer, Dordrecht

\bibitem[\protect\citeauthoryear{Bertin}{Bertin}{1997}]{bertin97}
Bertin G.,  1997, ApJ, 478, L71

\bibitem[\protect\citeauthoryear{Bertin}{Bertin}{2000}]{bertinbook2}
Bertin G.,  2000, Dynamics of Galaxies.
Cambridge University Press, Cambridge

\bibitem[\protect\citeauthoryear{Bertin \& Lodato}{Bertin \&
  Lodato}{1999}]{BL99}
Bertin G.,  Lodato G.,  1999, A\&A, 350, 694

\bibitem[\protect\citeauthoryear{Bertin \& Romeo}{Bertin \&
  Romeo}{1988}]{bertinromeo88}
Bertin G.,  Romeo A.~B.,  1988, A\&A, 195, 105

\bibitem[\protect\citeauthoryear{Boss}{Boss}{2002}]{boss02}
Boss A.~P.,  2002, ApJ, 576, 462

\bibitem[\protect\citeauthoryear{Collin \& Zahn}{Collin \&
  Zahn}{1999}]{collin99}
Collin S.,  Zahn J.~P.,  1999, A\&A, 344, 433

\bibitem[\protect\citeauthoryear{Coppi}{Coppi}{1980}]{coppi80}
Coppi B.,  1980, Comments on Plasma Physics and Controlled Fusion, 5, 261

\bibitem[\protect\citeauthoryear{Gallimore et~al.,}{Gallimore
  et~al.}{1997}]{gallimore97}
Gallimore J.~F.,  et~al., 1997, Nature, 388, 852

\bibitem[\protect\citeauthoryear{Gammie}{Gammie}{1996}]{gammie96}
Gammie C.~F.,  1996, ApJ, 457, 355

\bibitem[\protect\citeauthoryear{Gammie}{Gammie}{2001}]{gammie01}
Gammie C.~F.,  2001, ApJ, 553, 174

\bibitem[\protect\citeauthoryear{Goodman \& Tan}{Goodman \&
  Tan}{2003}]{goodman03}
Goodman J.,  Tan J.~C.,  2003, submitted to ApJ: \texttt{astro-ph/0307361}

\bibitem[\protect\citeauthoryear{Greenhill et~al.,}{Greenhill
  et~al.}{1996}]{greenhill96}
Greenhill L.~J.,  et~al., 1996, ApJ, 472, 21

\bibitem[\protect\citeauthoryear{Greenhill \& Gwinn}{Greenhill \&
  Gwinn}{1997}]{greenhill97}
Greenhill L.~J.,  Gwinn C.~R.,  1997, Astrophysics \& Space Science, 248, 261

\bibitem[\protect\citeauthoryear{Hohl}{Hohl}{1971}]{hohl71}
Hohl F.,  1971, ApJ, 168, 343

\bibitem[\protect\citeauthoryear{Johnson \& Gammie}{Johnson \&
  Gammie}{2003}]{gammie03}
Johnson B.~M.,  Gammie C.~F.,  2003, ApJ, 597, 131

\bibitem[\protect\citeauthoryear{Laughlin \& Bodenheimer}{Laughlin \&
  Bodenheimer}{1994}]{laughlin94}
Laughlin G.,  Bodenheimer P.,  1994, ApJ, 436, 335

\bibitem[\protect\citeauthoryear{Laughlin \& R\`o\.zyczka}{Laughlin \&
  R\`o\.zyczka}{1996}]{laughlin96}
Laughlin G.,  R\`o\.zyczka M.,  1996, ApJ, 456, 279

\bibitem[\protect\citeauthoryear{Launhardt \& Sargent}{Launhardt \&
  Sargent}{2001}]{launhardt2001}
Launhardt R.,  Sargent A.~I.,  2001, ApJ, 562, 173

\bibitem[\protect\citeauthoryear{Levin}{Levin}{2003}]{levin03}
Levin Y.,  2003, preprint: \texttt{astro-ph/0307084}

\bibitem[\protect\citeauthoryear{Lin \& Pringle}{Lin \& Pringle}{1987}]{lin87}
Lin D. N.~C.,  Pringle J.~E.,  1987, MNRAS, 225, 607

\bibitem[\protect\citeauthoryear{Lodato \& Bertin}{Lodato \&
  Bertin}{2001}]{LB2001}
Lodato G.,  Bertin G.,  2001, A\&A, 375, 455

\bibitem[\protect\citeauthoryear{Lodato \& Bertin}{Lodato \&
  Bertin}{2003}]{LB03a}
Lodato G.,  Bertin G.,  2003, A\&A, 398, 517

\bibitem[\protect\citeauthoryear{Lynden-Bell \& Kalnajs}{Lynden-Bell \&
  Kalnajs}{1972}]{lyndenbell72}
Lynden-Bell D.,  Kalnajs A.~J.,  1972, MNRAS, 157, 1

\bibitem[\protect\citeauthoryear{Lynden-Bell \& Pringle}{Lynden-Bell \&
  Pringle}{1974}]{lyndenbell74}
Lynden-Bell D.,  Pringle J.~E.,  1974, MNRAS, 168, 603

\bibitem[\protect\citeauthoryear{McCaughrean \& O'Dell}{McCaughrean \&
  O'Dell}{1996}]{mccaughrean96}
McCaughrean M.~J.,  O'Dell C.~R.,  1996, AJ, 111, 1977

\bibitem[\protect\citeauthoryear{Mayer et~al.,}{Mayer  et~al.}{2002}]{mayer02}
Mayer L.,  et~al., 2002, Science, 298, 1756

\bibitem[\protect\citeauthoryear{Monaghan}{Monaghan}{1992}]{monaghan92}
Monaghan J.~J.,  1992, ARA\&A, 30, 543

\bibitem[\protect\citeauthoryear{Murray}{Murray}{1996}]{murray96}
Murray J.~R.,  1996, MNRAS, 279, 402

\bibitem[\protect\citeauthoryear{Navarro \& Steinmetz}{Navarro \&
  Steinmetz}{1997}]{navarro97}
Navarro J.,  Steinmetz N.,  1997, ApJ, 478, 13

\bibitem[\protect\citeauthoryear{Nelson et~al.,}{Nelson
  et~al.}{2000}]{nelson2000}
Nelson A.~F.,  et~al., 2000, ApJ, 529, 357

\bibitem[\protect\citeauthoryear{Paczy\'nski}{Paczy\'nski}{1978}]{pacinski78}
Paczy\'nski B.,  1978, Acta Astronomica, 28, 91

\bibitem[\protect\citeauthoryear{Pickett et~al.,}{Pickett
  et~al.}{2000}]{pickett2000}
Pickett B.~K.,  et~al., 2000, ApJ, 529, 1034

\bibitem[\protect\citeauthoryear{Pringle}{Pringle}{1981}]{pringle81}
Pringle J.~E.,  1981, ARA\&A, 19, 137

\bibitem[\protect\citeauthoryear{Rice, Armitage, Bate \& Bonnell}{Rice
  et~al.}{2003a}]{rice03a}
Rice W. K.~M.,  Armitage P.~J.,  Bate M.~R.,    Bonnell I.~A.,  2003a, MNRAS,
  338, 227

\bibitem[\protect\citeauthoryear{Rice, Armitage, Bate \& Bonnell}{Rice
  et~al.}{2003b}]{rice03b}
Rice W. K.~M.,  Armitage P.~J.,  Bate M.~R.,    Bonnell I.~A.,  2003b, MNRAS,
  339, 1025

\bibitem[\protect\citeauthoryear{Sandell \& Weintraub}{Sandell \&
  Weintraub}{2001}]{sandell2001}
Sandell G.,  Weintraub D.,  2001, ApJ, 134, 115

\bibitem[\protect\citeauthoryear{Shakura \& Sunyaev}{Shakura \&
  Sunyaev}{1973}]{shakura73}
Shakura N.~I.,  Sunyaev R.~A.,  1973, A\&A, 24, 337

\bibitem[\protect\citeauthoryear{Shu, Tremaine, Adams \& Ruden}{Shu
  et~al.}{1990}]{shu90}
Shu F.,  Tremaine S.,  Adams F.~C.,    Ruden S.~P.,  1990, ApJ, 358, 495

\bibitem[\protect\citeauthoryear{Sirko \& Goodman}{Sirko \&
  Goodman}{2003}]{sirko03}
Sirko E.,  Goodman J.,  2003, MNRAS, 341, 501

\bibitem[\protect\citeauthoryear{Thacker, Tittley, Pearce, Couchman \&
  Thomas}{Thacker et~al.}{2000}]{thacker00}
Thacker R.~J.,  Tittley E.~R.,  Pearce F.~R.,  Couchman H. M.~P.,    Thomas
  P.~A.,  2000, MNRAS, 319, 619

\bibitem[\protect\citeauthoryear{Toomre}{Toomre}{1964}]{toomre64}
Toomre A.,  1964, ApJ, 139, 1217

\end{thebibliography}

\appendix

\section{Angular momentum transport induced by artificial viscosity}

\begin{figure*}
\centerline{\psfig{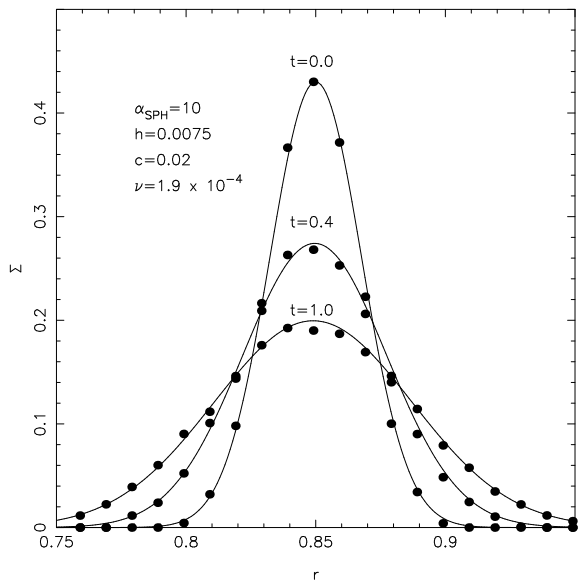}}
\caption{Radial surface density plots at times of $t=0.0$, $t=0.4$,
and $t=1.0$ for an axisymmetric ring evolved using SPH with the
standard artificial viscosity (dots) and evolved analytically (solid
lines) with a viscosity determined using the parameters of the SPH
calculation.}
\label{fig:ringspread1}
\end{figure*}

\begin{figure*}
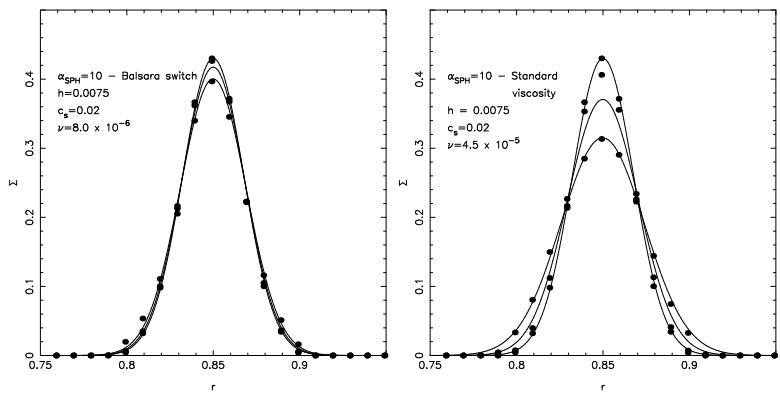

\centerline{\psfig{figure=ringspreadbals.ps,width=2.0truein}
            \psfig{figure=ringspreadnorm.ps,width=2.0truein}}
\caption{Radial surface density plots at times of $t=0.0$, $t=0.4$,
and $t=1.0$ for an axisymmetric ring evolved using SPH (dots) and
evolved analytically (solid lines).  In the left hand figure the
Balsara switch was used in the SPH artificial viscosity while in the
right hand figure the simulation was evolved using the standard SPH
viscosity.  The Balsara switch clearly reduces the effective shear
viscosity by a factor of between $5$ and $10$.}
\label{fig:ringspread2}
\end{figure*}

In this Appendix we discuss the magnitude of the angular momentum
transport associated with the artificial SPH viscosity, in order to be
sure that the main contribution to the shear stress described in
Section \ref{sec:results} is actually due to the effect of
gravitational disturbances.

\subsection{The Balsara switch}

As mentioned in the main text, we have adopted the Balsara form of
viscosity in order to reduce the effect of artificial viscosity in
transporting angular momentum. Although this modification to the
standard SPH viscosity has been studied in some detail
\citep{navarro97,thacker00}, it would be useful to have some idea of
the reduction in shear viscosity that occurs when using this form of
the artificial viscosity.

In the continuum limit, the linear term in the standard SPH artificial
viscosity can be shown to have the following form \citep{lubow94,murray96}
\begin{equation}
\nu = \frac{1}{8}\alpha_{\mathrm{SPH}} c_{\mathrm s} h
\label{eq:visc}
\end{equation}
where $\alpha_{\mathrm{SPH}}$ is the linear viscosity coefficient,
$c_{\mathrm s}$ is the sound speed and $h$ is the SPH smoothing length
(which essentially determines the resolution of the simulation). As
shown by \citet{lyndenbell74}, the time evolution of the surface
density and radial velocity of an initial Gaussian ring can be
determined analytically. To test the standard SPH viscosity,
\citet{murray96} therefore considered the evolution of an initial
Gaussian ring with the initial radial velocity determined
analytically, using Eq. (\ref{eq:visc}) to determine the viscosity. We
have repeated this calculation and show the results in Figure
\ref{fig:ringspread1}. The SPH calculation had a linear viscosity
coefficient of $\alpha_{\mathrm{SPH}} = 10$, a sound speed of
$c_{\mathrm{s}} = 0.02$, and a constant smoothing length of $h =
0.0075$. The choice of the SPH parameter was made in order to be
consistent with \citet{murray96}. As in \citet{murray96}, the pressure
forces were switched off so as to study the artificial viscosity in
isolation. This, on the one hand, prevents the ring from spreading due
to pressure forces, but on the other hand would cause the disc to
collapse to the midplane.  As a check that converging flows do not
influence our results, we have also performed some tests with a fixed
vertical coordinate of the SPH particles, and we found no significant
differences. The dots in Figure \ref{fig:ringspread1} show the time
evolution of the SPH surface density.  For the chosen SPH parameters,
the associated shear viscosity, according to Eq. (\ref{eq:visc}), is
$\nu = 1.9 \times 10^{-4}$. The solid lines in Figure
\ref{fig:ringspread1} show the analytic viscous evolution of an
initial Gaussian ring with an imposed shear viscosity of $\nu = 1.9
\times 10^{-4}$. As in \citet{murray96} the SPH evolution of the
initial Gaussian ring follows the analytic result very closely and
Eq. (\ref{eq:visc}) appears to be a good representation of the shear
viscosity associated with the linear term in the standard SPH
artificial viscosity.

To perform a similar calculation to determine the viscous transport
when using the Balsara switch is more subtle since the shear viscosity
should, ideally, go to zero.  This will, of course, not be exactly
true in practice, but we cannot now determine what the initial radial
velocity profile should be. To get some idea of the viscous transport
when using the Balsara switch we have considered the time evolution of
a Gaussian ring in which, since ideally we expect no shear viscosity,
we set the initial radial velocities to zero. For linear viscosity
coefficients of $\alpha_{\mathrm{SPH}} = 0.1$ and
$\alpha_{\mathrm{SPH}} = 1$ there is no noticeable spreading between
$t = 0$ and $t = 1$.  For $\alpha_{\mathrm{SPH}} = 10$, however, the
ring did spread slightly.  The left hand side of Figure
\ref{fig:ringspread2} shows the SPH evolution of an initial Gaussian
ring (dots) plotted at times of, as in Figure \ref{fig:ringspread1},
$t = 0.0$, $t = 0.4$, and $t = 1$ for $\alpha_{\mathrm{SPH}} = 10$.  A
direct comparison with the analytical prediction for the surface
density evolution and with the results shown in
Fig. \ref{fig:ringspread1} would be misleading in this case, because
initially the ring has to settle down before attaining the appropriate
radial velocities resulting from the viscous evolution. As a result,
the initial evolution of the ring is slower than predicted
analytically by \citet{lyndenbell74}. In the left panel of Figure
\ref{fig:ringspread2} we also plot the analytic evolution of an
initial Gaussian ring (solid line) with an imposed viscosity of $\nu =
8.0 \times 10^{-6}$, which appears to best describe the average
evolution of the ring.  In order to compare these results with those
obtained from the use of the standard SPH viscosity, we performed an
identical calculation (with the initial radial velocities set to zero)
except using the standard SPH viscosity without the Balsara
switch. This is shown in the right hand side of Figure
\ref{fig:ringspread2}. Again the dots represent the evolution of the
SPH surface density at times of $t=0.0$, $t=0.4$, and $t=1$ and the
solid line shows the analytic evolution of the surface density. Also
in this case, as discussed above, the initial evolution is somewhat
slower than expected. The average evolution between $t=0$ and $t=1$
can be fitted with a viscosity $\nu = 4.5 \times 10^{-5}$. These
results, though not conclusive, indicate that the Balsara switch
reduces the effective shear viscosity by a factor of between $5$ and
$10$. By using the Balsara switch we should therefore be able to
reduce the angular momentum transport due to the artificial shear
viscosity without, according to \citet{thacker00}, significantly
influencing the handling of shocks.

A comparison between Eq. (\ref{eq:visc}) and
Eq. (\ref{eq:alphashakura}) readily shows that the artificial shear
viscosity in SPH (not using the Balsara swith) can be described in
terms of the $\alpha$ parametrisation in the following way:
\begin{equation}
\label{eq:alphaartificial}
\alpha_{\mathrm{art}}=\frac{1}{8}\alpha_{\mathrm{SPH}}\frac{h}{H}.
\end{equation}
Eq. (\ref{eq:alphaartificial}) therefore shows that the magnitude of
$\alpha_{\mathrm{art}}$ depends on how well the vertical structure of
the disc is resolved. In most of our simulations we typically had
$H\approx 5h$, so that, if we had used the normal SPH viscosity, we
would expect $\alpha_{\mathrm{art}}\approx 2.5\times 10^{-3}$, for
$\alpha_{\mathrm{SPH}}=0.1$ (the value used in all our
simulations). The use of the Balsara switch enables us to further
reduce this contribution to $\alpha_{\mathrm{art}}\approx 5\times
10^{-4}$.

\begin{figure}
\centerline{\psfig{figure=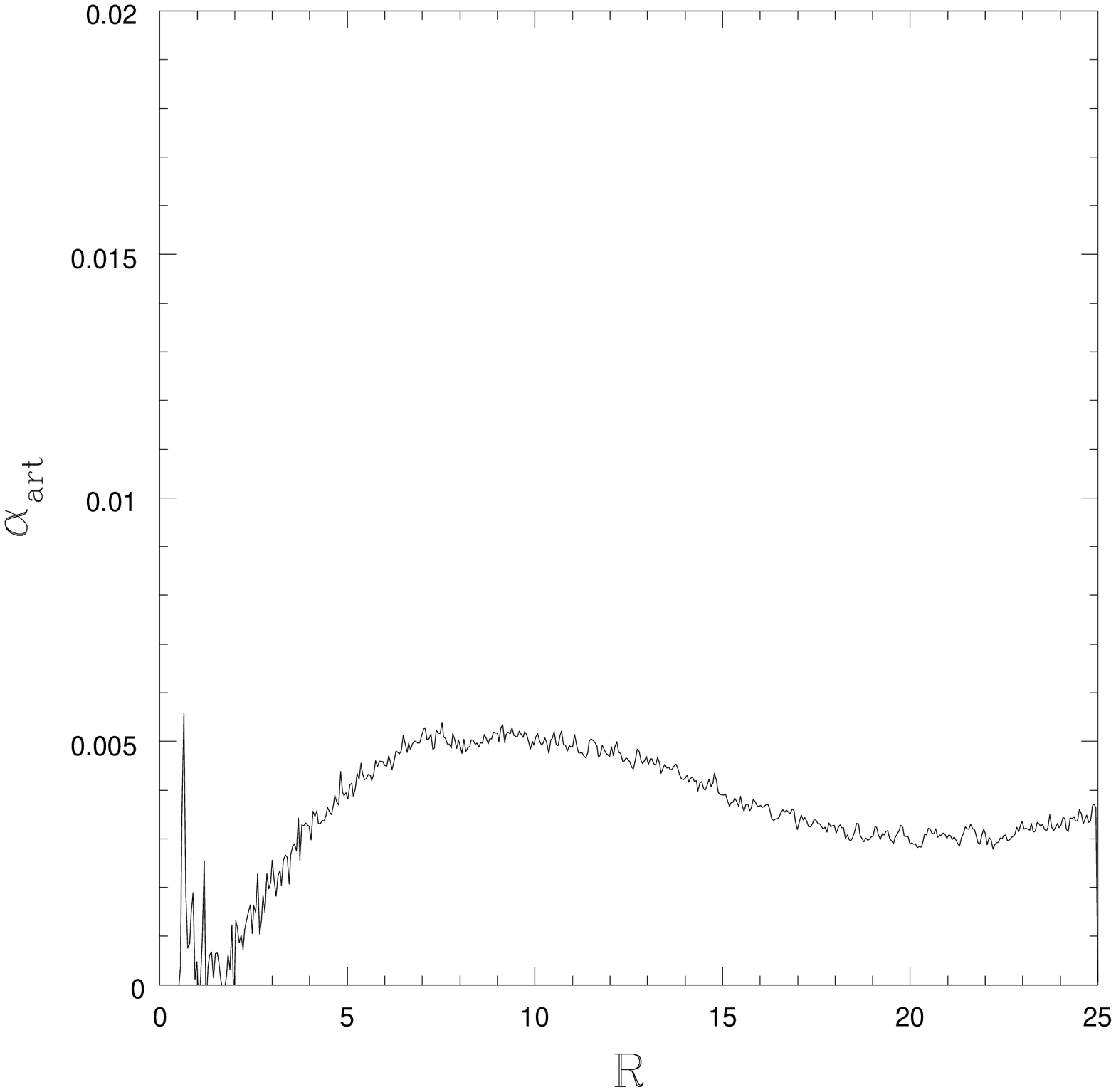,width=84mm}}
\caption{Angular momentum transport induced by artificial viscosity in
our SPH simulations, as parametrised by an effective
$\alpha_{\mathrm{art}}$.}
\label{fig:test}
\end{figure}

\subsection{Artificial transport in disc simulations}

As discussed in the previous Section, thanks to the Balsara switch, we
expect the linear term in the viscosity to produce only a minor
contribution to the shear stress. However, the quadratic term of the
artificial viscosity (parameterised by $\beta_{\mathrm{SPH}}$) might
still give some contribution to the angular momentum transport. In
addition, in our self-gravitating disc simulations, we have also added
pressure force, which might as well contribute to the stress, as found
also by \citet{murray96}. Moreover, we would also like to test this
issue in a less idealised case, with respect to the simple spreading
ring calculations described in the previous Section.  For this
purpose, we have performed some simulations taking the same surface
density profile as in Section \ref{sec:results} (i.e. $\Sigma\propto
R^{-1}$) and a constant $Q$ profile, with $Q=1$, but in which we have
switched off the main source of transport and dissipation (i.e. the
disc self-gravity) and the cooling. The artificial viscosity
coefficients were $\alpha_{\mathrm{SPH}}=0.1$ and
$\beta_{\mathrm{SPH}}=0.2$, as in the simulations presented in Section
\ref{sec:results}.  We have computed the Reynolds stress according to
Eq. (\ref{hydro}), and obtained an effective $\alpha_{\mathrm{art}}$
from Eq. (\ref{alpha}). The results are shown in
Fig. \ref{fig:test}. The magnitude of this artificial transport is
never larger than $\alpha_{\mathrm{art}}\approx 5~10^{-3}$.  Since our
results (see Section \ref{sec:results}) indicate that the transport
induced by gravitational torques can be parametrised with an effective
$\alpha\approx 5\times 10^{-2}$, we can be confident that the major
contribution to the stress tensor comes from gravitational torques
rather than from artificial viscosity.

\end{document}